\documentclass[10pt]{article}  
\usepackage[letterpaper, left=.75in, top=.75in, right=.75in, bottom=.7in,nohead,includefoot]{geometry}
\usepackage{charter,graphicx,amsfonts,amsmath,amssymb,latexsym,placeins,enumitem,natbib} 
\usepackage[svgnames,x11names]{xcolor}

  \def\blu{\color{RoyalBlue3}}	
\def\eqn#1{eqn.~\eqref{eq:#1}}
\def\pyi{p_i(y_i)}\def\Pyi{P_i(y_i)}
\def\diag{\textrm{diag}}

\newcommand{\seq}[2]{#1\,{:}\,#2}
\def\para#1{\smallskip\noindent{\em #1}.}

\def\c{\mathbf{c}}

\def\f{\mathbf{f}} 
\def\m{\mathbf{m}}
\def\q{\mathbf{q}}

\def\y{\mathbf{y}}

\def\one{\mathbf{1}}
\def\A{\mathbf{A}}

\def\F{\mathbf{F}}

\def\V{\mathbf{V}} 
\def\zero{\mathbf{0}}
\def\one{\mathbf{1}}
\def\diag{\textrm{diag}} 
\def\optf{\f^*}  \def\optfi{f_i^*}  \def\optlambda{\lambda^*}  \def\optfone{f_1^*}        \def\optfn{f_n^*}

\def\bgamma{\boldsymbol{\gamma}}

\def\today{November 2021} 

\begin{document}  
\setcounter{page}{0}\thispagestyle{empty}

\begin{center} 
{\bf \Large Perspectives on Constrained Forecasting} 

\bigskip
{\blu{\large{\bf Mike West}} }

\medskip
Department of Statistical Science, Duke University, Durham NC 27708-0251, U.S.A.

\medskip
{\blu{mike.west@duke.edu} }

\bigskip
{\em Revised version}: \today
\end{center}

\begin{center}{\bf Abstract} \end{center} 

This expository paper discusses Bayesian decision analysis perspectives on problems of constrained forecasting. 
Foundational and pedagogic discussion contrasts decision analytic approaches with the traditional, but typically inappropriate,  inferential approach. 
Illustrative examples include development of novel constrained point forecasting and entropic tilting methodology to explore consistency of a predictive distribution with an imposed or hypothesized constraint. 
Linear, aggregate constraints define illuminating examples that relate to broadly important problems involving  aggregate and hierarchical constraints in commercial and economic forecasting.   Discussion explores the impact of different loss functions,   questions of how constrained forecasting is impacted by dependencies among outcomes being predicted, and promotes the broader use of decision analysis including routine evaluation of 
predictive distributions of loss under chosen forecasts/decisions.  
Extensions to more general constrained forecasting problems, connections with broader interests in forecast reconciliation and other considerations are noted.

 
\vfill
 
\small 
\para{Acknowledgements}  
Research reported here was motivated by applied forecasting problems with researchers in the Science Team at  84.51$^\circ$, 100 West 5th Street, Cincinnati, OH 45202, and benefited in particular from  discussions with 
Andrew Cron, Paul Helman and Christoph Hellmayr.  Discussions with Domenico Giannoni of Amazon, and with current and recent graduate students in Statistical Science at Duke University including Lindsay Berry, Daniel Deng, Isaac Lavine and Anna Yanchenko, contributed further perspectives. 
The author also acknowledges constructive comments and suggestions from other readers. 
\normalsize
 
\newpage

\section{Introduction \label{sec:introduction}} 
 
 There are many areas in which it is of interest to assess a constraint, or set of constraints, against a posterior or predictive distribution.  The goals may include conditioning a distribution on constraints as well as understanding compatibility of constraints with a given distribution.   This paper discusses this general question using simple examples; the perspective is foundational and pedagogic.  The area is open for new research and methodology development, and this is anticipated with discussion of decision theoretic alternatives to traditional probabilistic approaches that are typically inappropriate. 
 
The examples focus on constrained forecasting in which a constraint is to be explored and/or imposed on a given predictive distribution. 
Motivating settings are in commercial and economic systems where problems of aggregate and constrained forecasting are common and important. Broad issues  include consistency of forecast models, distributions and point forecast selection at different levels of aggregation in time and/or dimension,  and  conditioning forecasts for sets of series on those of others, often higher-level aggregates. The Bayesian forecasting literature has a long history in these areas,  but  core challenges remain and are increasingly important in large-scale forecasting with many intersecting levels of aggregation.  Roles for Bayesian decision analytic approaches have been relatively unexploited; this paper highlights new opportunities for methodological and applied progress using decision analysis.

Among practitioners of Bayesian forecasting, conditioning forecast distributions on assumed values of totals or aggregates has been routine for decades, with early work going back to the 1970s in formal  models. Detailed discussion, with references, can be found 
in~\citet[][section~16.3]{West1997}.  Such approaches condition predictive distributions on assumed constraints, i.e., take a purely probabilistic-- or, more broadly inferential-- view that constraints are information to condition upon~\citep[e.g.][]{HarrisonGreen1973,deAlba1988,deAlba1992,deAlba1993,West1997}.  The increased interest in constrained and hierarchical forecasting with major scaling of data, time series and complex hierarchies has continued to build on these foundations. 
Beyond fully subjective Bayesian approaches,  the field has explored related Bayesian moment-based approaches~\citep{deAlba2006} as well as related non-Bayesian approaches~\citep{Guerrero1989,GuerreroNieto1999,Hyndman2019}. One main theme in such approaches is to exploit variants of constrained least-squares or linear Bayes'  methods~\citep{GoldsteinWooff2007}.  
Technically these approaches share features with the inferential results in constrained 
multivariate normal  distributions, while the perspective is again that of adjusting inferences in estimation/prediction settings.

There are both foundational and technical challenges with the solely inferential approach.  A first and main point is that probabilistic conditioning is generally not justified in settings where constraints are imposed as a matter of intervention.    Implicit in the probabilistic approach is routine Bayesian learning under which realized constraint values-- totals or other aggregates-- arise as random draws from the underlying model distributions for outcome quantities to be constrained. This is not often the case in applications. There, assumed constraint values are often chosen to explore \lq\lq what-if''  implications, so are imposed externally on the initial forecast distribution. In other contexts they are taken as representative values from a different, external model with a view to understand implications on relevant forecast values for the initial context. Echoing~\citet{Lindley1992ValenciaIV}, this argues, in part, for a decision analytic view, at least as a complement to the traditional inferential view.

From a technical viewpoint, probabilistic conditioning of joint distributions on totals or other aggregates becomes challenging outside of normal or related least squares approaches. Increasingly prevalent contexts involve time series of  non-negative counts~\citep[e.g.][]{Chen2017,ChenETALdynets2016JASA,Soyer2018,ChenBanksWest2018,BerryWest2018DCMM,BerryWest2018TSM,West2020Akaike}. In such settings, 
normal/linear Bayes' approaches are inappropriate and, if applied, can  generate misleading results such as negative point forecasts and always non-integer values. Further  challenges arise with increasingly high-dimensional time series for which posited deterministic constraints are increasingly likely to represent outcome regions that are \lq\lq rare events'' (i.e., out in the tails) of joint forecast distributions. This raises foundational and practical questions of how, when and whether to proceed to condition probability forecasts.  
Finally, predictive distributions are often represented as  Monte Carlo samples. Conditioning  Monte Carlo representations of joint distributions on  constraints can be addressed in various ways, such as with importance sampling or adaptive ABC-style approaches~\citep[][and references therein]{BonassiWestBA2015}. However, such approaches are,
inherently limited theoretically~\citep[e.g.][]{RussellEtAl2013} and very challenged in realistic applied contexts of even modest dimensions. 

With this background, the current paper explores conditioning using Bayesian decision analysis to contrast with the probabilistic approach.    This begins with discussion of optimal Bayesian point forecast selection subject to constraints. Beyond the basics of exploring the roles of different loss functions and the optimization questions to compute constrained point forecasts, this discussion emphasizes the broader pay-offs of a full Bayesian analysis that can come from exploring the full predictive distributions of outcome losses.  The second general decision analysis development involves entropic tilting (ET). This defines a nice methodological bridge between traditional probabilistic conditioning and the  developments of Bayesian point forecast and optimization approaches.

Section~\ref{sec:setting} discusses constrained forecasting in the simplest setting of overlaying a sum constraint on an outcome vector to be predicted. With examples, this section discusses the traditional probabilistic approach, Bayesian decision analysis perspectives to define constraint point forecasts, and  introduces the Bayesian entropic tilting approach. 
Section~\ref{sec:decisionanalysisexamples} details relevant loss functions,  highlighting some of central use and importance in commercial and economic forecasting with positive outcomes, with additional technical development in the Supplementary Material.
Section~\ref{sec:examples} discusses  examples using multivariate lognormal forecast distributions, including a 100$-$dimensional example related to applications in constrained commercial forecasting. 
Section~\ref{sec:conclusion} concludes with general comments and discussion of extensions. 
Supporting technical details and additional illustrative examples are given in Supplementary Material.

\section{Total Constrained Forecasting \label{sec:setting}} 

\subsection{Setting and Background} 

At one time point in a forecasting analysis of a set of $n$ series, the outcome of interest $\y=(y_1,\ldots,y_n)'$ has predictive distribution $P(\y)$ with margins $\Pyi, \ (i=\seq1n).$  The corresponding p.d.f.s are $p(\y)$ and $\pyi$ whether the distributions are discrete, continuous or mixed;  the density (p.d.f.) terminology is used with the 
corresponding general Stieltjes notation for expectations with the understanding that this covers all cases. Key interests are in discrete time series including binary and non-negative counts as well as in more traditional contexts where the $y_i$ are continuous and often positive.   

Total constrained forecasting of $\y$ is the simplest but  most important example context of linear constrained analysis. This conveys the foundational concepts and issues, provides access to some analytically tractable examples that generate insights, and forms the basis of more general cases based on sets of constraints, including hierarchical constraints. In an hierarchical context, conditioning forecasts of the $y_i$ may be defined via a cascade in which the forecasts of $Y$ are themselves based on higher-level totals or other aggregates, with obvious recursive extension to multi-level hierarchies.

Let $Y=\one'\y = \sum_{i=\seq1n} y_i$  with implied distribution $P(Y)$ and p.d.f. $p(Y).$  Interest lies in forecasting $\y$ given a specific value $F$ for this total.  The value $F$ may be a chosen point forecast, such as $E(Y)=F$ under $P(\cdot),$   a point forecast generated from some external model or source,  or a \lq\lq what-if?'' value from a set being explored to understand impact of potential total constraints on the $y_i.$  It may also be just one value generated from an external or alternative model for $Y$ alone, representing one of a set of Monte Carlo draws against which interest lies in understanding implications for $\y$ based on $p(\y)$ coupled with the  information defined by that external source. 

\subsection{Traditional Inferential Perspective: Probabilistic Conditioning}

The traditional Bayesian/probabilistic view is that conditioning on the value of $F$ is a purely inferential question, and that it implies modifying $P(\y)$ to $P(\y|F)$, the conditional distribution  of $\y$ given $Y=F.$  The implied analysis identifies $P(\y|Y)$ for any total $Y,$ and then plugs-in the value $Y=F$
~\citep[][section~16.3]{West1997}.

\subsubsection{Theoretical Examples} 

{\em Examples 1: Normal, and Other Elliptically Symmetric Cases.}  If $P(\y)$ is given by $\y \sim N(\m,\V)$, then $Y \sim N(M,q)$ where $M=\one'\m$ and $q=\one'\c$ where $\c$ is the covariance vector $\c \equiv C(\y,Y)
= \V\one.$  It follows that  $P(\y|F)$ is (singular) normal with  mean $ \m_F= \m + \c(F-M)/q$ and (singular) variance matrix $ \V-\c\c'/q$. 

 If $P(\y)$ is a multivariate T, or other elliptically symmetric distribution, the location of $P(\y|F)$ is modified   as in the normal case, while dispersion depends on $F$ and increases in $|M-F|.$  The construction of  elliptically symmetric distribution as normal scale mixtures defines the underlying probability calculus and interpretation. For example, $\y\sim T_k(\m,\V)$,
implies  $\y|F \sim T_k(\m_F,\V_F)$-- now singular $T_k$-- with $\m_F$ as above and
$\V_F=(\V+\c\c'/q)v_F $ where $v_F = \{k+(F-M)^2)/q\}/(k+n).$   Uncertainty in $P(\y|F)$ naturally inflates as a function of the lack of concordance of  the value of $F$ with $p(Y)$.   
 
 \para{Examples 2: Lognormal and Log-T Cases}  
Major areas of application in business and economic analysis use conditionally normal, linear models-- such as dynamic linear models (DLMs:~\citealp{West1997,PradoFerreiraWest2021})-- for log transformed data. Implied predictive distributions are  log-T distributions on the original $\y$ data scale.  Normal approximations may be used, but are inadequate in contexts of restricted T degrees of freedom such as arise routinely, for example, in multivariate volatility modelling~(e.g.,~\citealp{PradoFerreiraWest2021}, chapter 10, and~\citealp{West2020Akaike}, section 2). 
The challenge then is that $p(Y)$ and $p(\y|Y)$ are not available analytically, so raising difficult questions of computation.  This is a severe constraint generally, but particularly when interest lies in fast and scalable analysis to accurately evaluate aspects of $p(\y|F). $   

\para{Examples 3: Discrete Cases}   Similar comments apply to distributions arising in increasingly large-scale models for discrete time series, including binary and non-negative count data~\citep{BerryWest2018DCMM,BerryWest2018TSM}.   Even in relatively simple models based on conditional Poisson forms for univariate series, dependencies across series destroy the ability to evaluate joint and conditional distributions analytically.

\subsubsection{Simulation-based Analysis} 
 
In many realistic models $P(\y)$ is represented via a Monte Carlo sample, so that evaluating and using $p(\y|Y)$ is a major challenge. Approaches such as adaptive importance sampling~\citep[e.g.][]{West1993} and   sequential Monte Carlo including
approximate Bayesian computation (ABC:~\citealp[e.g.][]{BonassiWestBA2015}) can be considered. However,  such methods do not reliably deal in any generality with the problem of conditioning on totals, or other aggregates, in problems of practicable dimension.  The problem has been considered in other areas too, and shown to be a very major challenge as well as NP-hard in discrete contexts~\citep{RussellEtAl2013}; it stands as an open challenge  to computational statistics. Again, applied contexts increasingly require fast, reliable and scalable analysis, which is simply not (yet) available.  

In low-dimensional problems, a vanilla ABC-style accept/reject method can sometimes prove useful.     Such an approach defines a subspace $S$ of the sample space of $\y$ consistent with values of $Y=\one'\y$ \lq\lq near'' $F$.  Simulated values $\y\sim P(\y)$ are then accepted if and only if  $Y\in S.$ 
The example in Section~\ref{sec:examples}  takes $ S = \{ \y:   (1-\tau)F \le Y \le (1+\tau) F \}$ for positive, small $\tau$ so that   
the maximum percentage error $100 \tau$ defines \lq\lq nearness'' of simulated $Y$ values to the target $F$.   
Accepted values are then approximately distributed as $P(\y|F)$ with approximation based on $\tau.$ The acceptance probability $p_S = Pr(\y\in S)$ under $P(\y)$ 
can be trivially estimated from the Monte Carlo samples and gives a guide to how effective the approach is given any value of $\tau.$

\subsection{Decision-Guided Probabilistic Conditioning: Entropic Tilting \label{sec:ETtheory} }

Imposing constraints in  forecasting is often essentially not an inference problem.  
Asking questions about how to forecast $\y$ given the total $Y=F$ imposed from an external model or source moves outside the formal probability model;  the imposed value of $Y=F$, or a collection of values to consider, did not arise from the $p(Y)$ implied by $p(\y).$  The value of $F$ is imposed by intervention, so that  asking about how fixing $Y=F$ should impact forecasts for $\y$ is   more naturally a decision question. At the least, exploring decision analysis perspectives is an opportunity to broaden the framework and examine approaches complementary to the traditional, probabilistic view.   Entropic tilting (ET) is one fully Bayesian decision analysis approach.

\subsubsection{Entropic Tilting and Moment Constraints}

With respect to the baseline distribution $P(\y)$,  ET is a maximum entropy related approach that aims to choose a distribution $G(\y$)
as a modification of $P(\y)$ that satisfies a set of expectation constraints 
$E_g[\q(\y)]=\zero$ where $\q(\y)$ is a $q-$vector of functions of $\y$.    
 An example with $q=2$ takes  $\q(\y) = (Y-F, Y^2-F^2-V)'$  
with  $Y=\one'\y$  and some specified $F$ and $V>0;$  then  $G(\y)$ implies a distribution for $Y$ that   has mean $F$ and variance $S$ (relevant only, of course, in contexts where variances exist).    
ET is an explicit decision analytic approach that aims to select $G(\y)$ \lq\lq close'' to $P(\y)$ subject to the expectation constraints;  
ET defines \lq\lq close''  in terms of the Kullback-Leibler divergence (KLD) of $P(\y)$ from $G(\y).$  With p.d.f.s $p(\y)$ and $g(\y),$  the solution 
is the exponentially-tilted form $g(\y)  = c_{\bgamma} \exp\{\bgamma'\q(\y)\}  p(\y)$ where the $q-$vector $\bgamma$ explicitly enforces the expectation constraints and $c_{\bgamma}>0$ is the normalizing constant.    The KLD optimal $\bgamma$ can typically be numerically computed using a Newton-Raphson (NR), algorithm to solve the $q-$vector equation $\int_{\y}\q(\y) \exp\{\bgamma'\q(\y)\} dP(\y) =\zero$, with NR utilizing  the second derivative matrix 
$\int_{\y} \q(\y)\q(\y)' \exp\{\bgamma'\q(\y)\}  dP(\y).$ 
In introducing ET to the econometrics community, one of the key contributions of~\cite{RobertsonET2005}  was to note that Bayesian analysis of $P(\y)$ is often/typically based on Monte Carlo simulation; this makes the integrals in such optimization computations accessible via direct Monte Carlo approximation.   

\subsubsection{Novel Development of ET for Conditioning on Constraints}   
ET can be used  to define $G(\y)$ to approximately conform with deterministic constraints.  Focus now explicitly  on the context of a single constraint (whether linear or non-linear), so
 $q=1$,  $\q(\y)=q(\y)$ is a scalar function and $\bgamma=\gamma$ is scalar. 
Take the specific choice $q(\y) = I( \y\in S) - (1-\epsilon)$ for some subspace $S$ of the sample space of $\y$,
indicator function $I(\cdot)$ and a specified probability $\epsilon$. Applying ET  then yields a distribution $G(\y)$ under which $Pr(\y_i\in S)=1-\epsilon$. This is very general and can be used, for example, to map $P(\y)$ to a $G(\y)$ having a specified median.    Applying this to approximate the constrained problem of the current paper involves taking: 
(i)  the subspace $S$ very small and concentrated around a constrained value of a deterministic function of $\y$; and (ii) the tolerance $\epsilon$ small, i.e.,
$1\gg\epsilon>0.$    The resulting  $G(\y)$  approximately satisfies the constraint with level of approximation defined by how small $S$ and $\epsilon$ are.  
The example in Section~\ref{sec:examples} for the context with $Y=\one'\y$ 
takes $ S = \{ \y:   (1-\tau)F \le Y \le (1+\tau) F \}$  where $1\gg\tau>0;$   here $\tau$  defines concentration of $S$ based on percentage error 
of $Y$ from $F$ being no more than $100 \tau$.

In this very specific ET context, the solution $G(\y)$ can be directly evaluated without resort to numerical optimization.  Note that the p.d.f is 
$g(\y) \propto \exp\{\gamma I( \y\in S) \} p(\y)$, or 
$g(\y) = c \{ \exp(\gamma) I(\y\in S) + 1-I(\y\in S) \} p(\y)$ where $c$ is the normalizing constant; clearly,
$c^{-1} = \exp(\gamma) p_s + 1-p_s$ where $p_s=Pr(\y\in S)$ under $P(\y).$          Then, since the expectation of $q(\y) $ under $G(\cdot)$ is constrained to be $1-\epsilon,$ the optimal $\gamma$ is the solution to  $1-\epsilon =c \exp(\gamma) p_s;$  this is easily solved to give 
$\gamma=\log\{(1-\epsilon)(1-p_s)/(\epsilon p_s)\}.$   Note that this is decreasing in both $\epsilon$ and $p_s,$  and will generate $\exp(\gamma)\gg 1$ when $\epsilon$ and/or $\tau$ are small, consistent with an increasingly binding constraint. 

Considering cases when $P(\y)$ is represented as a Monte Carlo sample, this  ties intimately with the ABC-style accept/reject approach noted above.
Suppose, with no loss of generality, that the Monte Carlo sample is equally weighted. 
The ABC analysis leads to a weighted sample in which a sampled vector $\y\sim P(\y)$ is given a weight of 1 if $\y\in S,$ 0 otherwise. The ET analysis defines weights 
proportional to $\exp(\gamma)$ for samples $\y\in S,$ and proportional to 1 otherwise.   As noted, $\gamma$ will tend to be large in practice, so that this ET reweighting is close to accept/reject.   
It can further be shown that, with a large Monte Carlo sample size,  the effective sample size of the normalized ET weights converges around 
$c^{-2} / \{  \exp(2\gamma) p_s + 1-p_s\} $; in practical cases with $\gamma>>1,$ this is approximately $p_s,$ the acceptance probability of the ABC-style analysis. 

\subsection{Point Forecast Decision Analysis Perspective \label{sec:BDA}}
  
More traditional Bayesian decision analysis approaches focus on optimal point forecast selection under $P(\y)$ but now subject to the imposed constraints. 

Let  $L(\y,\f)$ be a loss function chosen to score a point forecast vector $\f$ of outcome $\y$. Standard Bayesian decision analysis chooses that point forecast vector $\optf$ that minimizes the expected loss subject to the constraints. That is, 
\begin{equation} \label{eq:constrainedoptimization} 
\optf = \textrm{argmin$_\f$ } R(\f)\quad \textrm{subject to} \quad \one'\f=F, \quad \textrm{where } R(\f)  = \int_\y  L(\y,\f) dP(\y). \end{equation}
The following section concerns technical developments using specific loss functions.  Some general comments on loss function structure and the broader applied perspective on Bayesian decision analysis are first noted. 

\subsubsection{Additive Losses} 
In many applied problems,  loss functions will be additive over  outcomes, i.e., $L(\f,\y) = \sum_{i=\seq1n}L_i(y_i,f_i)$ for individual loss functions 
$L_i(\cdot,\cdot)$ in each dimension. In forecasting sales or demand for sets of items $i,$ for example, the translation to revenue (gained or lost) per item is a primary consideration. Other contexts might extend to loss functions reflecting cross-item scores.  Development below focuses on  additive loss functions, leaving extensions to the reader and future, customized applications. 

\subsubsection{Generalizations with Multiple Constraints} 
The broader class of problems for hierarchical  and other sets of constraints simply extends the above formulation to involve the constraint
  $\A'\f = \F$  where $\A$ is a specified  $n\times k$ matrix of full rank $k<n,$  and $\F$ is a specified $k-$vector.  Analysis then targets minimization of 
  $R(\f)$ subject to these $k$ constraints.  For example,  constraints on sets of intersecting subtotals can be defined by a matrix $\A$ of zeros and ones, while other, more general weighted averages are obvious extensions.  

\subsubsection{Broader View: Distributions of Loss} 
Section~\ref{sec:introduction} has already raised the central question and potential importance  of exploring predicted loss distributions, i.e., considering aspects of $P(L(\y,\f))$ implied under $P(\y)$ for $\f=\optf$ (and possibly other values, perhaps \lq\lq close to'' $\optf$).   This perspective is one of evaluation and presentation of uncertainties in loss outcomes in decision analysis akin to the usual \lq\lq uncertainty quantification'' view in inference. Again, this simply argues for the broader view of decision analysis in exploring loss distributions, and this is a point of emphasis throughout this paper in the specific settings of constrained forecasting. 
 


\section{Decision Analysis and Classes of Loss Functions \label{sec:decisionanalysisexamples} }

\subsection{Lagrangian Formulation}
The  Lagrangian formulation for optimization in~\eqn{constrainedoptimization} is to choose $(\f,\lambda)$ to minimize 
\begin{equation} \label{eq:constrainedriskfunction} 
R(\f) + \lambda (F-\one'\f)
\end{equation}
with a real-valued Lagrange multiplier $\lambda$.     Assuming a minimizing solution $\optf(\lambda)$ given any allowable value of $\lambda,$ solving  $F=\one'\optf(\lambda)$ for $\optlambda$ defines the optimal forecast vector $\optf \equiv \optf(\optlambda) = (\optfone,\ldots,\optfn)'.$      

Details for specific loss functions  commonly used in forecasting applications are noted below (with  additional technical details in 
Supplementary Material). The standard use in unconstrained Bayesian decision analysis-- in  forecasting and parameter estimation-- is  background~\citep[e.g.][]{FrenchInsuaDTbook2010,JQSmithDA}. A main interest   is to present examples of optimal constrained forecasts for  such loss functions, and highlight differences and implications. As noted above, the development   uses an additive loss function $L(\f,\y) = \sum_{i=\seq1n}L_i(y_i,f_i)$ so that 
\begin{equation} \label{eq:additiveriskfunction} 
R(\f) = \sum_{i=\seq 1n} R_i(f_i)\quad\textrm{with}\quad R_i(f_i) = \int_{y_i} L_i(y_i,f_i) dP_i(y_i)dy_i, \ i=1:n,\end{equation}
where $P_i(y_i)$ is the marginal predictive  distribution of $y_i.$  The resulting $\optf$ does not  involve dependencies among the $y_i.$ However, it is critical for practical application to be aware that the resulting distributions of loss at the optimum (or at any other value of $\f$) are of course very much impacted by the joint structure of $P(\y)$, as examples in Section~\ref{sec:examples} below illustrate.  
 
In most practical contexts, there is no direct analytic solution to the implied optimization problem;  numerical methods are needed.    Assuming 
$\optf(\lambda)$ is available for any $\lambda,$ the optimal $\optlambda$ is solution to  $q(\lambda)=0$ where 
$ q(\lambda) = \one'\optf(\lambda) - F.$  
A direct Newton-Raphson (NR) algorithm is typically most efficient and effective in solving this, relying on the 
derivative function $\dot q(\cdot)$. The basis of NR iterations is as follows. 
 \begin{itemize}[noitemsep,topsep=0pt]%
\item {\em  Initialize:} Set iterate count $t=0$, and Lagrange multiplier value $\lambda = \lambda^0$, a chosen initial value; set
 $F^0= \one'\optf(\lambda^0).$
\item {\em  Iterate}:  For  steps $t\ge 1,$ compute   
$\lambda^t = \lambda^{t-1} - q(\lambda^{t-1})/\dot q(\lambda^{t-1}), $    then update 
the implied $\optf(\lambda^t)$ and the sum $F^t=\one'\optf(\lambda^t).$ 
\item {\em  Stop:} When changes in the sequence of scalars $\lambda^t$  and/or $|F-F^t|$ become \lq\lq small enough'', set $\optlambda = \lambda^t$ and $\optf = \optf(\optlambda),$ and stop.  
\end{itemize} 
Examples in Section~\ref{sec:examples} utilize this, with NR iterations typically converging very fast.  Depending on the chosen loss function,   the range of  values of $\lambda$ is restricted. This allows the analysis to be self-monitoring in that NR iterates moving $\lambda$ to a lower or upper bound  would indicate incompatibility of the conditioning value $F$ with the predictive distribution $P(Y).$ Such contexts are those in which enforcing the constraint might be questioned, and the algorithm will signal that. Finally, in contexts where predictions are based on Monte Carlo samples from $P(\y),$  implied Monte Carlo estimates will  be used to evaluate 
$\optf(\lambda)$ via direct, weighted or importance sampling.

\subsection{Squared Error Loss} 
Squared error (SE)  loss is  not of main applied interest in   commercial forecasting applications compared to other choices noted below. However, details are tractable and illuminating.   SE loss is, of course, restricted to models in which $P(\y)$ has finite second-order moments. Supposing this, 
let $m_i$ be the mean of $P_i(y_i)$ and $\m=(m_1,\ldots,m_n)'$ with sum $M=\one'\m.$ The $m_i$ are optimal unconstrained point forecasts under SE.   

Take  $L_i(y_i,f_i) = (y_i-f_i)^2/c_i$  where the $c_i>0$ can represent different scales or simply different weightings of forecast errors across the $n$ outcomes. 
Write $\c = (c_1,\ldots,c_n)'$ and $C=\one'\c.$  
Then simple quadratic optimization yields $\optfi(\lambda) = m_i + \lambda c_i/2$ for each $i=\seq 1n.$ Imposing the total constraint yields $\optlambda = 2(F-M)/C$ and thus 
 $\optfi = m_i + (F-M) c_i/C$.   These are the marginally optimal means $m_i$ corrected by the term $(F-M)c_i/C$;  this naturally represents an upward (downward) correction if $F$ exceeds (falls short of) the forecast mean of the total $E(Y) = M.$    While natural, it is clear that the scope for relevant application is proscribed; in addition to earlier comments on constraints for relevant applications, many applied interests concern integer, count, non-negative or bounded outcomes, and the inherent \lq\lq constrained least squares''  results lead to theoretically optimal forecasts that violate such inherent requirements.

\subsection{Absolute Deviation Loss} 
Absolute Deviation (AD) loss is perhaps the most important and widely used loss function in commercial forecasting as in other areas. 
Take  $L_i(y_i,f_i) = |y_i-f_i|/c_i$  where again $c_i>0$ are known weights.   Assuming finite first moments of the $P_i(y_i),$ 
it  follows that,  for any given $\lambda,$ \eqn{constrainedriskfunction} is minimized  over the $f_i$ at the values satisfying
$2P_i(f_i)-1 = c_i\lambda$ (see details in Supplementary Material). Thus 
$\optfi(\lambda) = P_i^-((1+\lambda c_i)/2)$ where $P_i^-(\cdot)$ is the inverse c.d.f. (quantile function) for each $i,$ whether discrete or continuous.  Note that the usual unconstrained forecast is the median of $P_i(y_i)$ in the case $\lambda=0.$   Otherwise,
  $\optfi(\lambda)$  is the $100 (1+\lambda c_i)/2$ percentile of $P_i(y_i)$. 
Note further  that $\lambda$ must lie in $[-r,r)$ where $r=1/\max_{i=\seq 1n} c_i$. 

\para{Example: Exponential Models} A purely illustrative, analytically tractable example highlights the analysis.  Suppose the $y_i$ are marginally exponential,  $y_i \sim Exp(1/m_i)$ with $m_i=E(y_i).$  The marginal medians are $\tilde f_i = m_i \log(2)$.  Take $c_i=1$ so $C=n$ and $\lambda \in [-1,1).$ It follows that $\optfi(\lambda) = m_i \log(2/(1-\lambda))$ for each $i,$ and imposing $F=\one'\optf(\lambda)$ yields $\optlambda = 1-2\exp(-F/M).$ As a result,  
the optimal forecast is  $\optf=\optf(\optlambda) = \m F/M.$   That is, each marginal mean $m_i$  is simply-- and very naturally-- scaled by the positive constant $F/M$  so that the resulting $\optfi  = m_i F/M$  sum to $F$. 
 
More generally, the direct Newton-Raphson (NR) algorithm for optimization  solves 
$q(\lambda)=0$ where $q(\cdot)$ and its derivative $\dot q(\cdot)$ are now given by  
$$ q(\lambda) = \sum_{i=\seq1n}  \optfi(\lambda) - F \quad\textrm{and} 
\quad\dot q(\lambda) =  \sum_{i=\seq1n}  c_i\{ 2 p_i(\optfi(\lambda)) \}^{-1}. $$ 
These are easily calculated when the marginal forecast distributions are of parametric forms, and via Monte Carlo approximations using forecast samples in other cases.

\subsection{Absolute Percent Error Loss and Variants} 
\subsubsection{APE Loss} 
For strictly positive outcomes $y_i>0,$ the modification of AD loss to a percent scale defines the absolute percent error (APE) loss that is simply key in commercial applications. APE puts forecast errors on a common scale (percent revenue, percent sales of numbers of items, etc.) so as to enable easy comparisons across outcomes and contexts~\citep[e.g.][]{BerryWest2018DCMM}. 
Assuming $P_i(\cdot)$ has support $y_i>0$ (perhaps bounded above), take  $L_i(y_i,f_i) = |y_i-f_i|/(y_ic_i)$ 
 where again $c_i>0$ are known weights.   Then the risk function component $R_i(f_i)$ for outcome $i$ has the form of the expected value of AD loss $|y_i-f_i|/c_i$ with respect to the modified distribution with density function $g_i(y_i)\propto p_i(y_i)/y_i.$    If this defines a p.d.f.,  then 
  $g_i(y_i) =k_i p_i(y_i)/y_i$ for some normalizing constant $k_i>0$ and    
 $$R_i(f_i) = c_i^{-1} \int_{y_i>0} |y_i-f_i| y_i^{-1} dP_i(y_i) = (c_ik_i)^{-1} \int_{y_i} |y_i-f_i| dG_i(y_i)$$
 where $G_i(\cdot)$ is the c.d.f. implied by p.d.f. $g_i(\cdot).$  
 Hence the AD analysis above applies with each $P_i(\cdot)$ replaced by $G_i(\cdot)$ and the weights $c_i$ replaced by $c_ik_i.$ 
 That is, theoretically and in the numerical evaluation using NR,  each  $\optfi(\lambda)$  is the $100 (1+\lambda c_i k_i)/2$ percentile of $G_i(y_i)$. 
 The following details and examples are to be noted. 
 \begin{itemize}[noitemsep,topsep=0pt]%
\item 
When $\lambda=0$ so that the constraint does not apply, the optimal forecasts $\optfi$ are the medians of the $G_i(\cdot)$, also known as the $(-1)-$medians of $P_i(\cdot).$   With typical positively skewed distributions on $y_i>0,$  these lie below the medians due to the greater mass at lower values under $G_i(\cdot)$ than under $P_i(\cdot).$  This feature is inherited in the constrained decision analysis as the relevant percentiles of $G_i(\cdot)$ for any given $\lambda$ will be similarly lower than those of $P_i(\cdot).$ 
\item 
Practical models include cases when the predictive distributions have forms related to those of compound shifted Poisson, compound gamma, lognormal and others.   As one theoretically tractable example revisited in Section~\ref{sec:examples} below,   suppose that  $P_i(\cdot)$ is lognormal,  $y_i \sim LN(m_i,v_i)$ with mode, median and mean of $y_i$ given by $\hat f_i = \exp(m_i-v_i),$   $\tilde f_i = \exp(m_i)$ and 
$\bar f_i = \exp(m_i+v_i/2), $ respectively.  It easily follows that 
$G_i(\cdot)$ is   $LN(m_i-v_i,v_i)$ and $k_i = \exp(m_i-v_i/2).$   Note that the $(-1)-$median of $P_i(y_i)$ is exactly its mode in this case. Percentiles of $G_i(\cdot)$ relevant in the constrained decision analysis solutions  can be very substantially smaller than those of $P_i(\cdot)$ when predictions are uncertain.
\item 
In contexts where predictions are based on Monte Carlo samples from $P(\y),$  implied Monte Carlo estimates of the percentiles of $G_i(\cdot)$ are easily evaluated using weighted or importance sampling.   
\item  In some cases, this analysis is infeasible as $p_i(y_i)/y_i$ is not integrable, whether available analytically or via simulation. Key cases with with real practical importance again include models generating log-T  predictive distributions; truncating the distributions to finite ranges is one modification enabling the analysis. 
\end{itemize}

\subsubsection{ZAPE Loss} 
In  discrete  cases when forecast distributions have non-zero probabilities on $y_i=0$ for some $i=\seq 1n,$ APE loss is not applicable.  Extension to   zero-adjusted absolute percent error (ZAPE) loss functions is then of interest~\citep{BerryWest2018DCMM,BerryWest2018TSM}. Suppose $y_i \ge 0$ and that the predictive distribution has a non-zero point mass 
$\pi_{i0} = P_i(0)$ at $y_i=0.$  A ZAPE loss function is 
$ L_i(y_i,f_i) = w_i(f_i) I(y_i=0)  +  |y_i-f_i|/(y_i c_i) I(y_i>0)$  where $w_i(f_i)>0$  penalizes point forecast $f_i$ when $y_i=0.$      \citet{BerryWest2018DCMM} show the relevance of ZAPE in forecasting sales of large numbers of consumer items when there are appreciable probabilities of \lq\lq no sales''.
In the current context, the   constrained APE analysis is easily extended; the emerging $\optfi(\lambda)$ may now include exact zero values for some outcomes $i$ across ranges of values of $\lambda.$   

For example, take $w_i(f_i)=f_i/c_i$ so that a point forecast $f_i=1$ when $y_i=0$ is penalized exactly as a point forecast $f_i=0$ when $y_i=1$~\citep{BerryWest2018DCMM}. 
Define the c.d.f.  $P_i^+(\cdot)$ for the c.d.f. $P_i(\cdot)$ constrained and renormalized on $y_i>0,$ with corresponding p.d.f. $p_i^+(y_i).$  Then
$$P_i(y_i) = \pi_{i0} I(y_i=0) + (1-\pi_{i0}) P_i^+(\cdot)I(y_i>0).$$  Then, define 
$G_i(\cdot)$ as the c.d.f. with p.d.f. $g_i(y_i)= k_i p_i^+(y_i)/y_i$ on $y_i>0$ where $k_i$ is the
with appropriate normalizing constant. 
With $w_i(f_i)=f_i/c_i,$ the risk component $R_i(f_i)$ satisfies
$$
c_i R_i(f_i) = \pi_{i0} f_i +  (1-\pi_{i0}) k_i^{-1}  \int_{y_i>0} |y_i-f_i| dG_i(y_i). 
$$
It  follows that,  for any given $\lambda,$ \eqn{constrainedriskfunction} is minimized over the $f_i$ at values given by 
$$
\optfi(\lambda) = \begin{cases} 0, & \textrm{if } u_i(\lambda)\le 0, \\  G_i^-(u_i(\lambda)), & \textrm{if } u_i(\lambda)> 0, \end{cases}
\quad\textrm{with}\quad  
u_i(\lambda) = \frac 12 \left\{1+ k_i \frac{(\lambda c_i-\pi_{i0})}{(1-\pi_{i0})}\right\},
$$
and where $G_i^-(\cdot)$ is the inverse of the c.d.f. $G_i(\cdot)$ (see additional details in Supplementary Material).
Here $\lambda$ must lie in $[s,r)$ with bounds given by 
$s=\max_{i=\seq 1n} \{((k_i+1)\pi_{i0}-1)/(c_ik_i)\}$ 
 and $r=\min_{i=\seq 1n} \{((k_i-1)\pi_{i0}+1)/(c_ik_i)\}.$   The results for APE are confirmed when $\pi_{i0}=0$ for all $i.$ 
Otherwise,  higher probabilities $\pi_{i0}$  will  lead to optimal point forecasts at zero. Extending to constrained forecasting is particularly interesting in such contexts. Technically, only minor modifications to the NR algorithm arise, with $q(\cdot)$ and its derivative now given by 
$$ q(\lambda) = \sum_{i=\seq1n} 
\optfi(\lambda)- F
 \quad\textrm{and}\quad 
 \dot q(\lambda) =  \sum_{i=\seq1n} I(u_i(\lambda)>0) c_ik_i \{ 2 g_i(\optfi(\lambda) \}^{-1}.$$ 
Practically important modifications of the above example of ZAPE include choices of the $w_i(f_i)$ penalty at $y_i=0$ that less heavily penalize 
larger values of the point forecasts $f_i$. In some contexts, the linear in $f_i$ penalty is  too dominant for larger values of $f_i$,  pushing the optimal $f_i^*$ to zero more aggressively than desired. In such settings, bounded weight functions such as $w_i(f_i) = f_i/(1+f_i)$ or 
$\min\{1,f_i\}$ are more relevant.    Examples 
using these forms can be easily implemented using 
extensions of the above optimization method, and bear out 
the effectiveness in reducing the overly aggressive shrinkage to zero of optimal forecasts while adding only modestly to computational load. 


\section{Illustrative Examples \label{sec:examples}}

\subsection{General Comments} 
As discussed above, some motivating applications involve  non-negative outcomes in commercial and allied areas.   Two illustrative examples reflect this, with  multivariate lognormal distributions that allow ranges of dependencies among the $y_i$. 
This setting provides access to some analytic tractability that aids in generating insights.  Related examples (not shown) using count data in which conditional Poisson models   linked via latent factors  share similar general features, though lack analytic tractability. 
The examples touch on differences in constrained point forecasts based on choice of loss function, and on how these vary with dependencies among the $y_i.$ They also focus on aspects of predictive distributions of losses as well as optimal point forecasts, a point stressed earlier that should always be part of the broader Bayesian decision analysis.

\subsection{Bivariate Lognormal Example \label{sec:bivarLNexamples} }

\subsubsection{Setting and Optimal Forecasts} 

\begin{figure}[p!]
\centering
\includegraphics[width=0.375\textwidth]{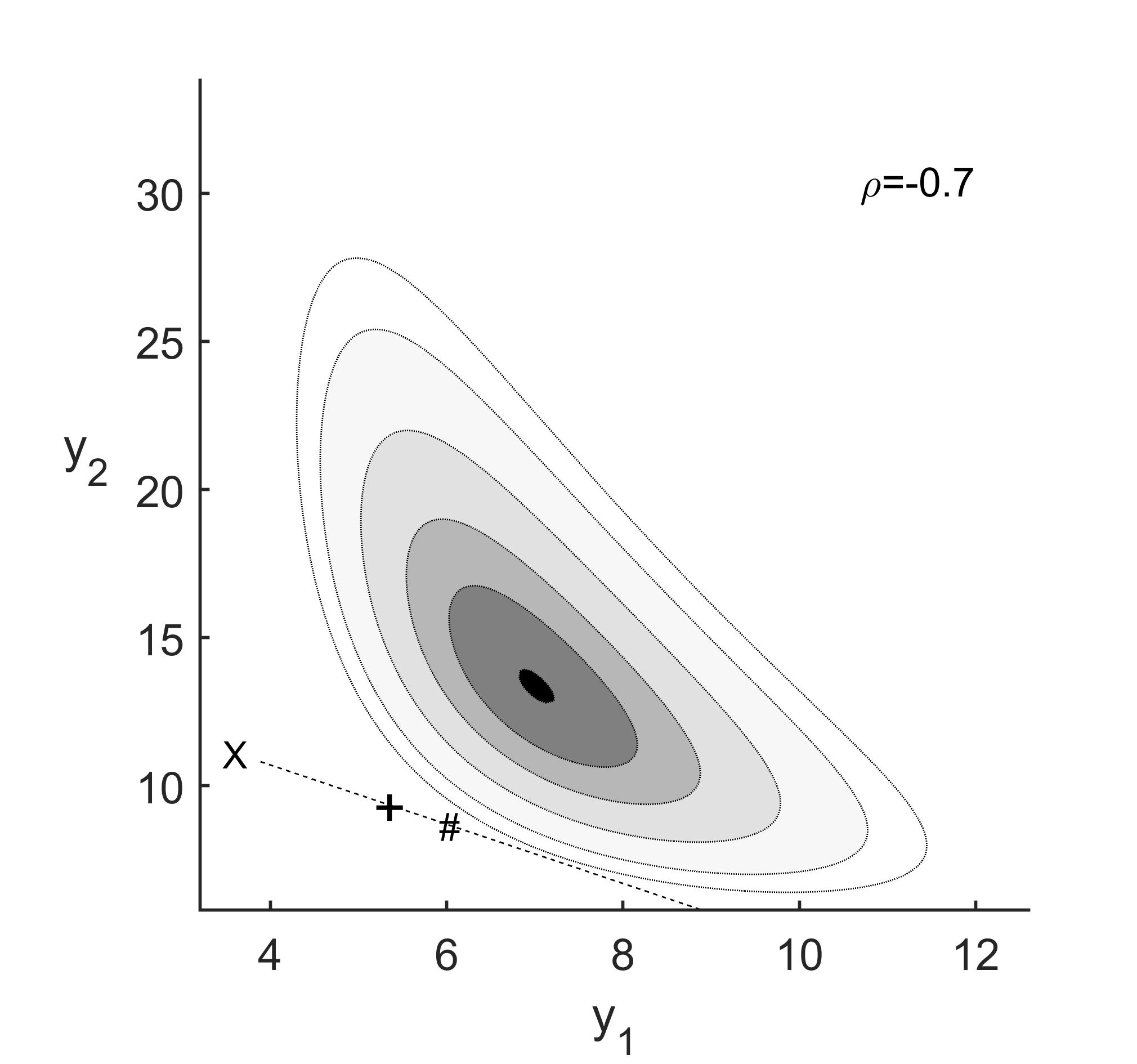}
\includegraphics[width=0.375\textwidth]{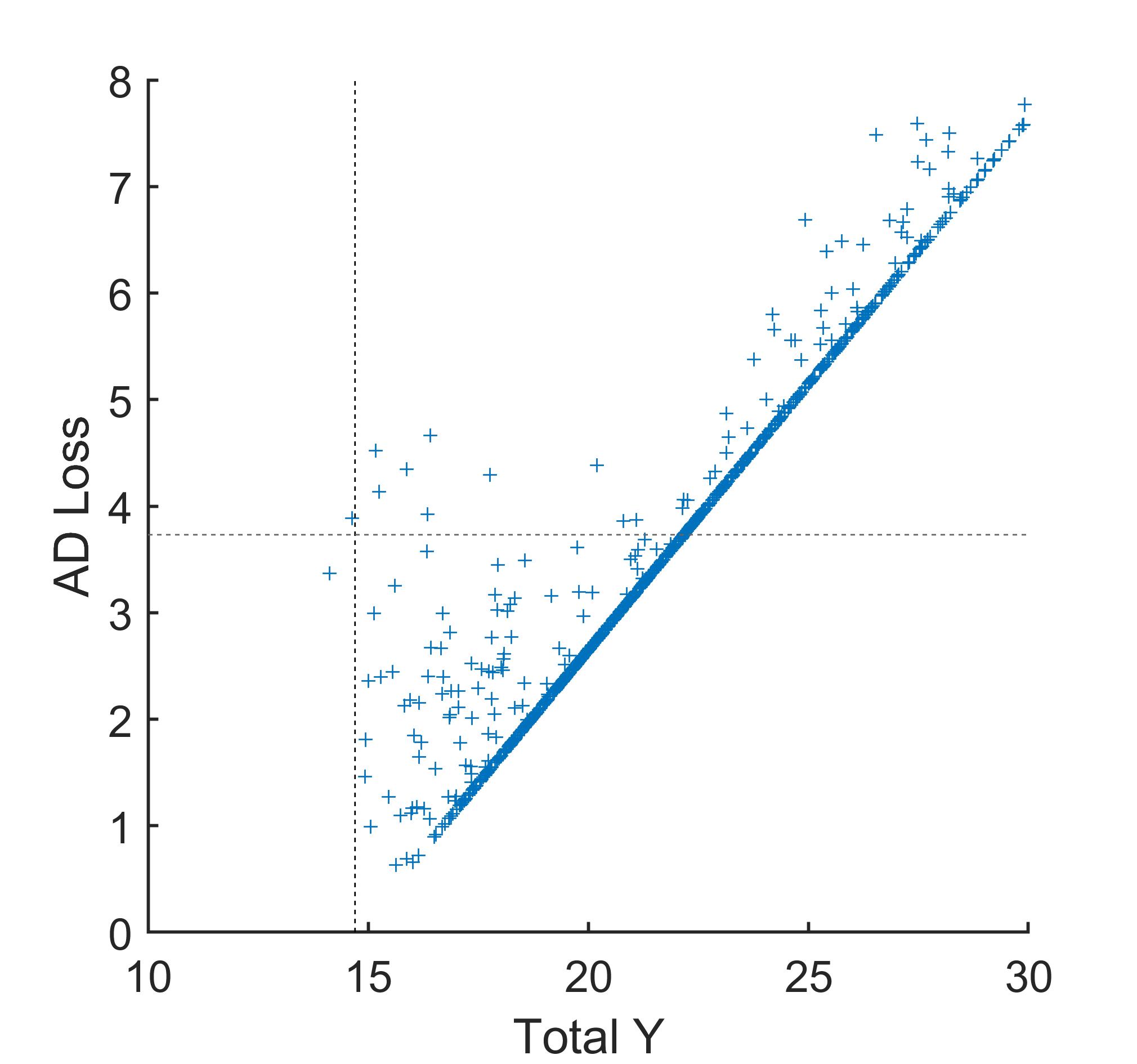}
\includegraphics[width=0.375\textwidth]{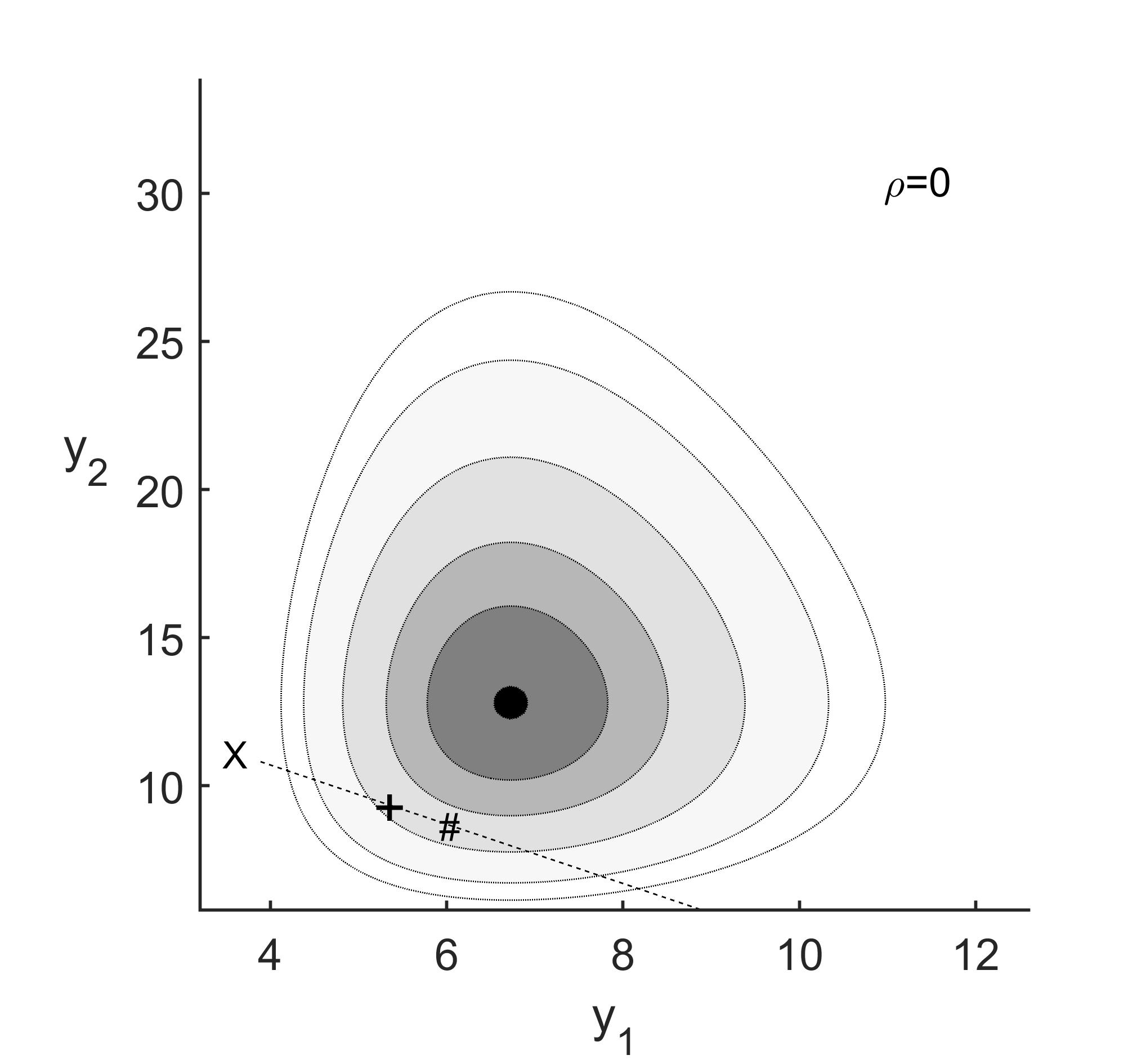}
\includegraphics[width=0.375\textwidth]{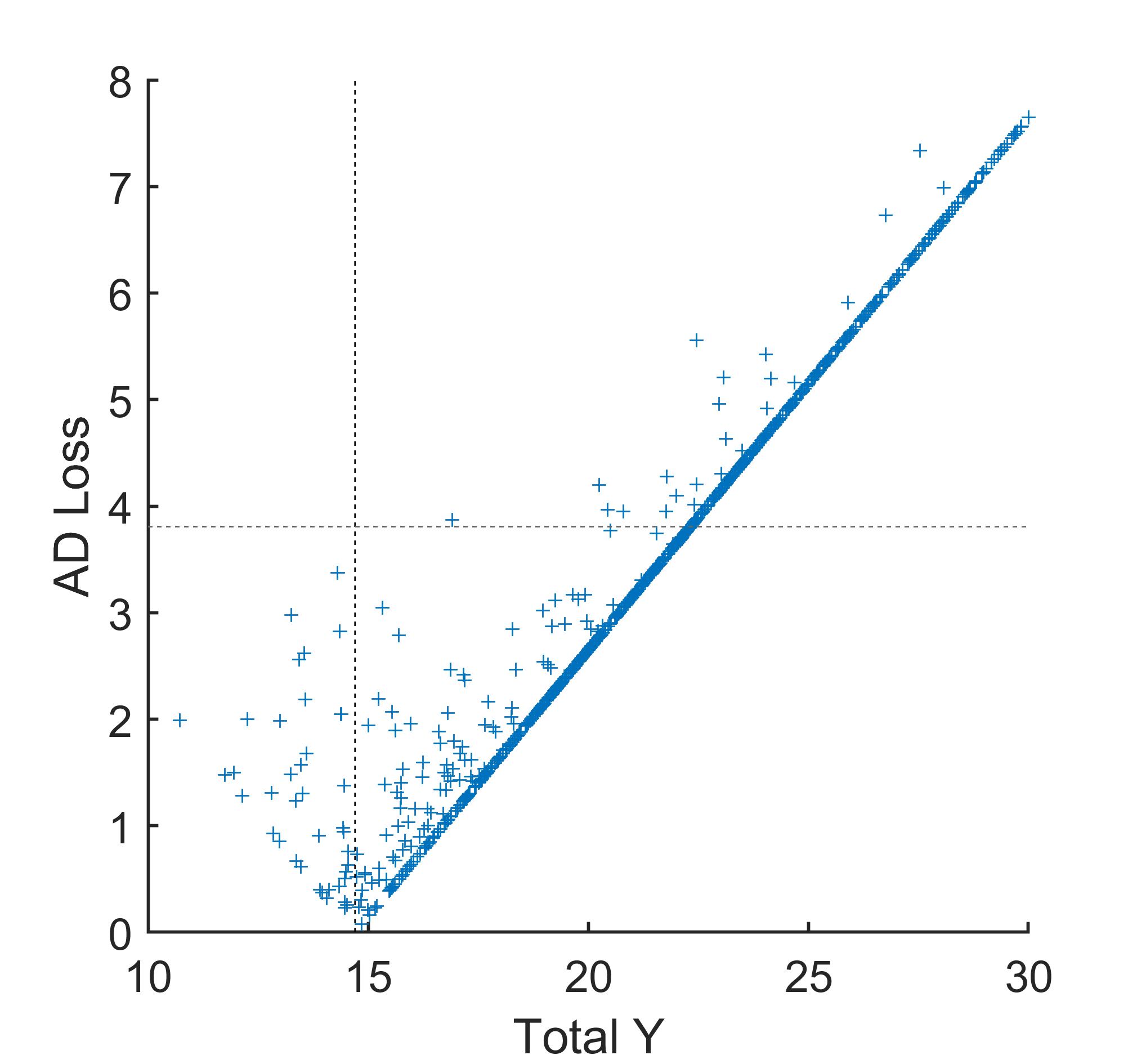}
\includegraphics[width=0.375\textwidth]{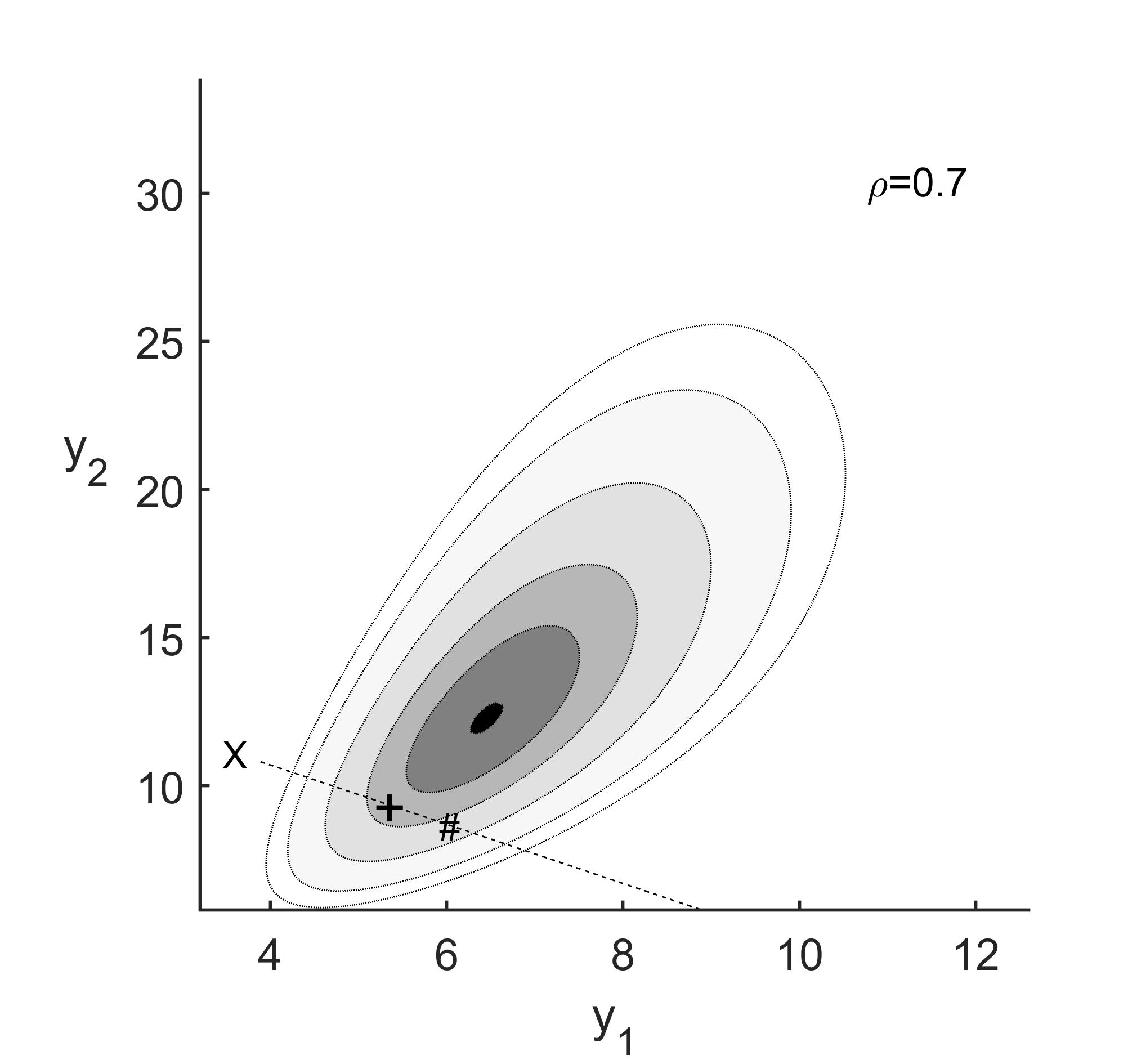}
\includegraphics[width=0.375\textwidth]{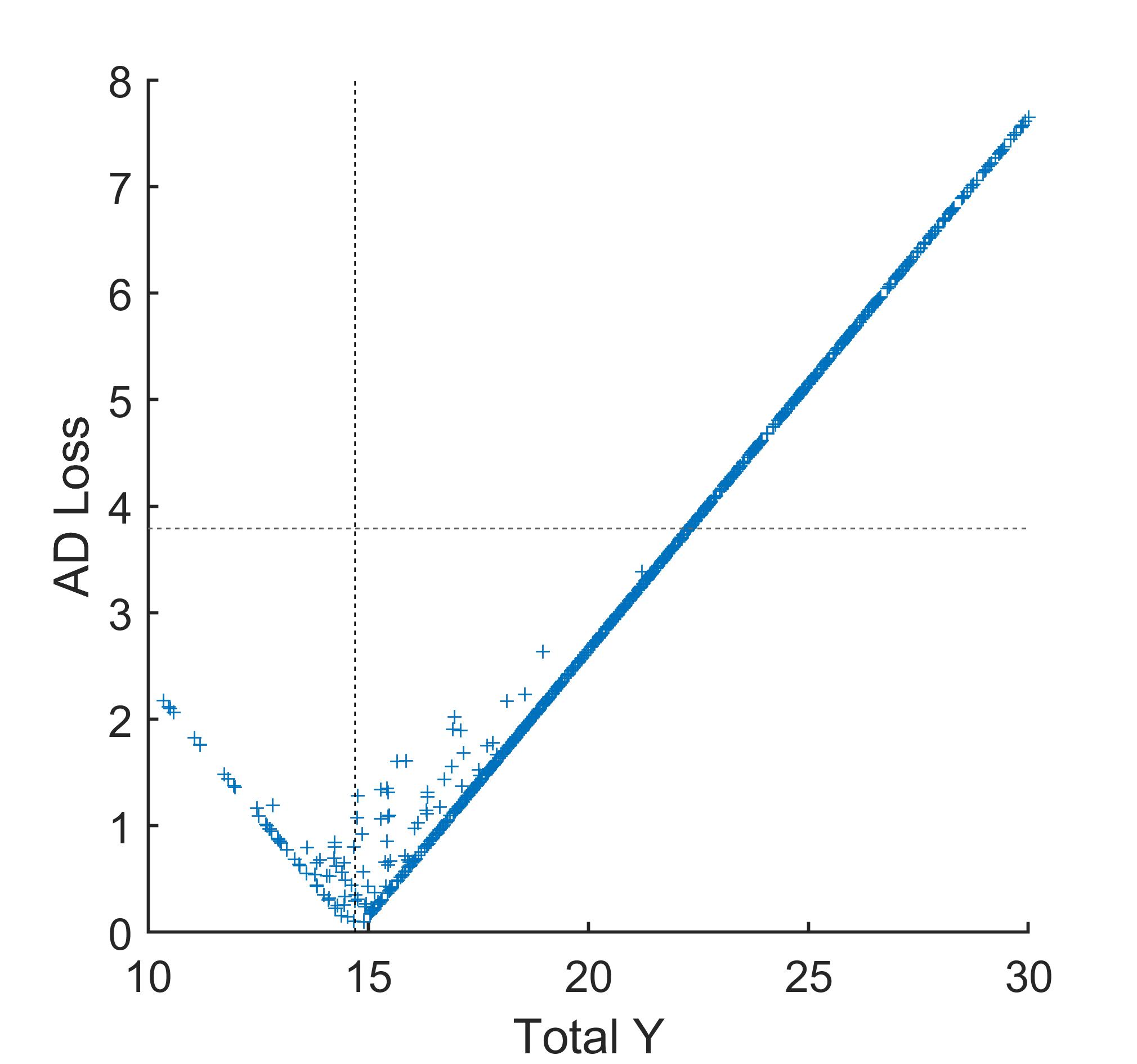}
\caption{Bivariate lognormal example with $F=14.7$, lying quite far into lower tail of $P(Y).$  {\em Left column:}  Contours of $p(\y)$ for three different levels of dependence $\rho\in \{-0.7,0,0.7\}$ and with the constraint $\one'\y=F$ indicated as the dashed line.  The symbols indicate the optimal point forecast vector $\optf$ under 
AD loss (+), APE loss (\#) and SE loss (X).  While marginal APE optimal forecasts are always lower than those under AD loss,  the joint constrained APE optimal forecast can be higher than AD optimal in some dimensions, simply due to the total constraint.   {\em Right column:}  Scatter plots of the corresponding joint predictive distributions of the outcome total and the {\em per dimension} loss at the value of $\optf$ i.e., a Monte Carlo sample from $P(Y,L(\y,\optf)/2).$  The vertical dashed line marks the value of the constraint, $Y=F$; the horizontal dashed line marks the value of the expected loss at the minimum, i.e., the optimized risk $R(\optf).$    
        \label{fig-1bivarLNcontoursandlossdstn} }
\end{figure}

A first set of examples has $n=2$ so that $\y'=(y_1,y_2).$  
The contours in Figure~\ref{fig-1bivarLNcontoursandlossdstn} are those of three bivariate lognormal distributions $\y\sim LN(\m,\V)$ 
whose parameters are the mean and variance matrix of the underlying bivariate normal for $(\log(y_1),\log(y_2))'.$ 
The examples have
$\m'=(\log(7),\log(14))$,  $\diag(\V)=(v_1,v_2)=(0.04, 0.09)$,  and the off-diagonal entry of $\V$ is $ 0.06 \rho$ 
for dependence parameter $\rho\in (-1,1).$  
The univariate lognormal margins $P_i(y_i)$ have
 modes-- that are also the $(-1)-$medians-- at  $\{6.73,12.80\}$, medians at $\{7,14\}$ and means at $\{7.14,14.64\}$.  
 The contours are those of the highest predictive density regions under $P(\y)$ with  $\{0.01,0.25,0.5,0.75,0.9,0.95\}$ probability content.  The three examples show contours for the cases of $\rho \in \{-0.7, 0, 0.7 \}.$ 
 
Under $P(\y),$ the sum of medians of the $y_i$ is 21, and the mean is $E(Y)=21.9.$   Figure~\ref{fig-1bivarLNcontoursandlossdstn} is based on $F=14.7,$ 
well into the lower tail of the forecast distribution $P(Y)$ so 
will lead to larger adjustments to constrained point forecasts.   The dashed lines define the total constraint, so  optimal point forecasts lie on these lines. 
The NR algorithm  converges in two or three steps to a high degree of precision. Searching for the AD optimal is initialized at marginal medians, and for the APE optimal at marginal $(-1)-$medians.  The figures show the values of the constrained AD, APE and SE point forecasts, showing differences  due to the choice of loss function.  While marginal $(-1)-$medians are always lower than marginal medians, the imposition of the constraint will often change the ordering in some dimensions.  Note also that the use of SE loss would be questioned in this context.   The Supplementary Material provides additional illustration with the same model but now 
using two other values of $F$-- one in the center of the predictive distribution $P(Y)$ and one in the upper tail-- with corresponding summaries.

\subsubsection{Loss Distributions} 
 
 Figure~\ref{fig-1bivarLNcontoursandlossdstn} also explores distributions of AD loss.     For each value of $\rho$,    Monte Carlo samples of $\y$ give samples from the joint distribution of $\{ Y, L(\y,\f) \}$ at any chosen $\f. $  The   figure shows resulting scatter plots at the AD-optimal $\f=\optf,$ with the simulated loss values scaled by $1/n=1/2$ so that the vertical axis is on a {\em per dimension} loss scale.    While $\optf$ is the same in all three cases, the distributions of optimized losses depend on the full joint $P(\y).$ The predictive distribution $p(Y)$ is more diffuse for positive values of $\rho$ than for zero or negative values, and this naturally translates into greater dispersion of the resulting distribution of losses. As $\rho$ varies in $\{-0.7,0,0.7\},$ the medians of the loss distributions are approximately 
$\{2.9, 2.7,2.5\}$, and the means are approximately $\{3.38, 3.18, 3.10 \};$ in each case, the minimum value of the risk function 
reduces as $\rho$ increases. However, the loss uncertainty increases as  $\rho$ increases; for example, the
 upper 95\% points of the loss distributions are approximately $\{ 6.86,7.45,7.90\}$ at these three values of $\rho.$ 
     Thus, while average or median risks define one order, the tail behaviour of  loss distributions raises additional considerations of possible \lq\lq downside'' losses.  
  Note also that there is appreciable probability on loss outcomes that are lower than the optimized risk, i.e., corresponding to the potential \lq\lq upside'' outcomes. 
 This argues for the broader view of decision analysis to understand aspects of loss distributions at optimal-- or other-- chosen point forecasts. It should be stressed that this is a general point-- not specific to constrained forecasting, but highlighted in this context.
 While the general concept 
 has been well-recognized in areas such as finance (with \lq\lq value-at-risk'' studies resulting) it is not generally appreciated in other areas of  decision analysis.


\subsubsection{Probabilistic Conditioning} 

\begin{figure}[hpbt!]
\centering
\begin{tabular}{cc}
\small Low  $F=14.7$ & \small High $F=24.15$ \\
\includegraphics[width=0.475\textwidth]{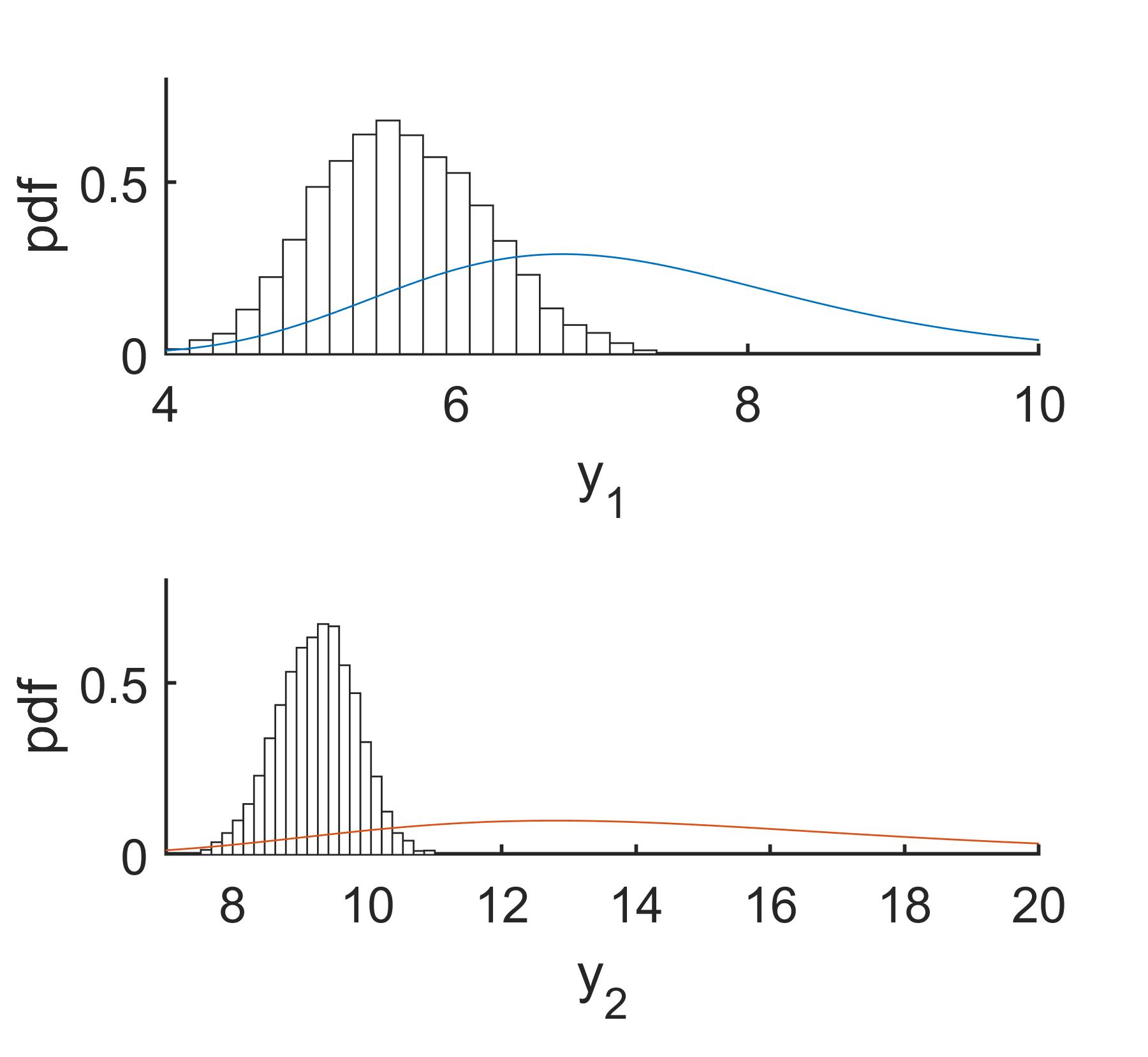} &
\includegraphics[width=0.475\textwidth]{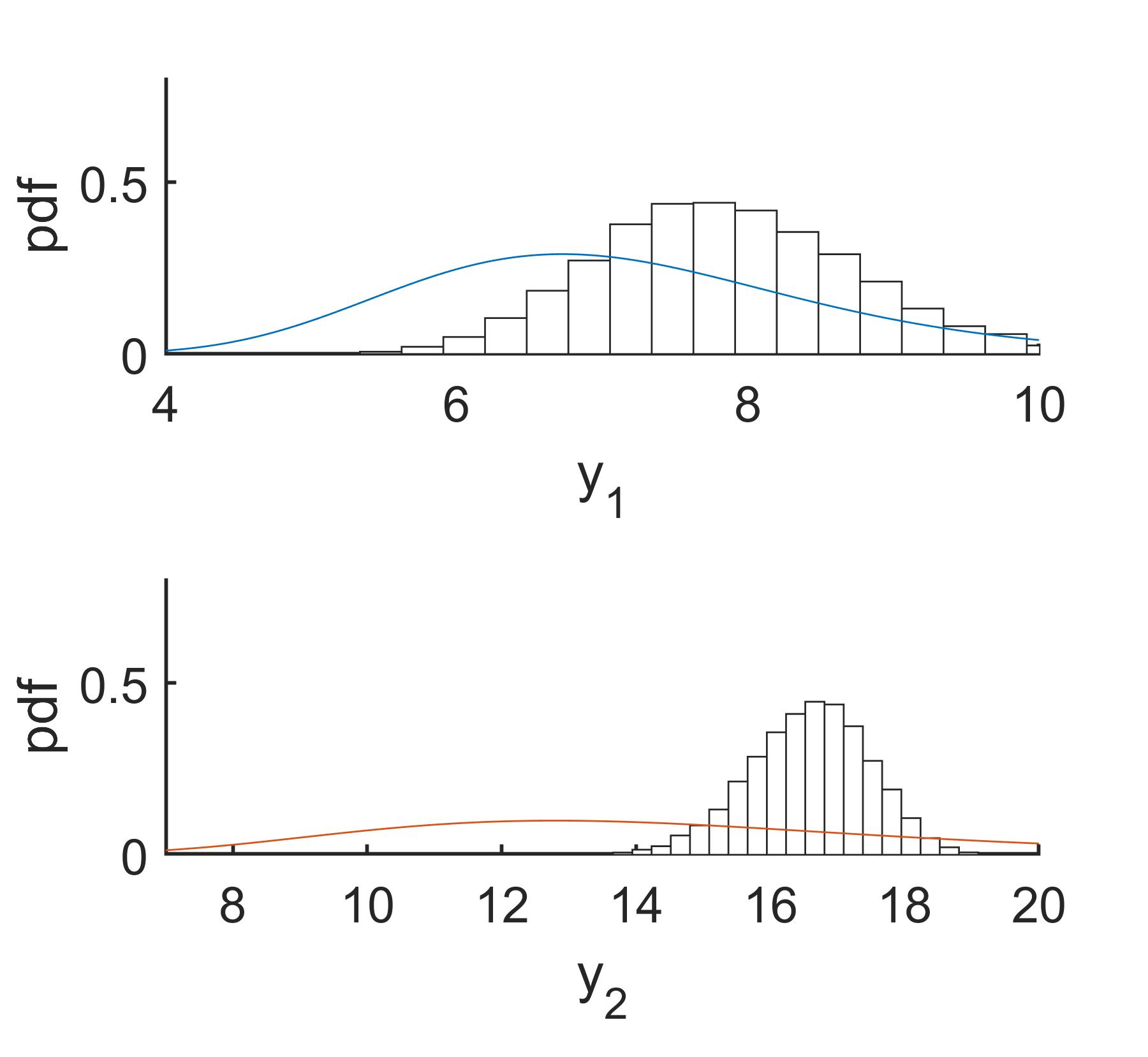} \\
\end{tabular} 
\caption{Bivariate lognormal example with $\rho=0.7$, showing marginal and approximate constrained p.d.f.s. Analysis generates a large Monte Carlo sample and accepts $\y$ if, and only if, the sum $Y$ satisfies $|Y-F|/F<  \tau$ with percent tolerance  $100\tau=0.5.$  The histograms represent ABC-approximate conditionals $p_i(y_i|Y=F)$ based on the joint prior $p(\y)$ whose lognormal margins $p_i(y_i)$ are displayed as curves. Analyses are based on the very low value of $F=14.7$ (left) and the high value of $F=24.15$ (right). 
	       \label{fig-bivarLNABC} }
\end{figure}

This example is a case in which the conditioning value of $F$ lies is well into the tails of $p(Y)$ which, as discussed earlier, represents challenges to the 
purely probabilistic approach of summarizing aspects of $p(\y|Y=F).$ That said, in this simple illustrative example in only $2-$dimensions, it is easy to generate very large Monte Carlo samples from $p(\y)$ and apply vanilla ABC-style methods.  Figure~\ref{fig-bivarLNABC} gives examples in the case of $\rho=0.7.$  
A large sample from $p(\y)$ was conditioned to simulated values of $\y$ such that $Y=\one'\y$ was \lq\lq close'' to the conditioning $F$ value, illustrating results for both the low and high values of $F$. Closeness was specified by 
 $|Y-F|/F<  \tau$ where $100\tau$  is the percent tolerance on this natural \lq\lq closeness to constraint'' metric.  Examples displayed use $\tau=0.005$; resulting scatter plots (not shown) of the constrained samples appear visually indistinguishable from the line $\one'\y=F.$   ABC acceptance rates at this tolerance are around 0.5-1.5\%, and smaller for negative values of $\rho$, indicative of the challenges of using probabilistic conditioning. In  realistic, higher-dimensional settings, this ABC-style analysis is simply not an option as (i) it becomes really challenging to define relevant tolerance ranges, and even with that in place (ii) 
the acceptance rates decay exponentially with dimension.   In contrast, the decision analysis approach has different goals, and generates useful and informative results of direct applied value in such contexts, and is at least complementary to the purely probabilistic approach. 
 
\subsubsection{Entropic Tilting} 

In the setting of Section~\ref{sec:ETtheory}, take ET function $q(\y) = I(y\in S) -(1-\epsilon)$ 
where the probability $\epsilon$ is close to zero and  $ S = \{ \y:   (1-\tau)F \le Y \le (1+\tau) F \}.$  This leads to ET optimal 
 $G(\y)\approx P(\y|F)$  with p.d.f. $g(\y) \propto \exp\{\gamma I( \y\in S) \} p(\y)$ with $\gamma =\log\{(1-\epsilon)(1-p_s)/(\epsilon p_s)\}.$   
In the current example, $P(\y)$ is defined in terms of a direct Monte Carlo sample $\y^i$ for $i=1:I$ where $I$ is the Monte Carlo sample size. These are reweighted according to weights  $w_i = \exp(\gamma)/\{ \exp(\gamma) $ for $ \y\in S,$ and $w_i \propto 1$ for $ \y\notin S,$ 
subject to summing to 1.  As noted earlier,  the optimal solution has $\gamma=\log\{(1-\epsilon)(1-p_s)/(\epsilon p_s)\}.$  
The optimal set of weights $w_i$ represent an importance sampling approximation to the optimal $G(\y)$  subject to the assessment of importance sampling accuracy using the usual metrics~\cite[e.g.][]{West1993}.  Inference on $\y$ conditional on the constraints then follows; one easy and oft-used step in importance sampling is to simply resample $\y$ from the set $\y^i$ according to the weights $w_i,$ and then proceed based on that resample (which, of course, will generally include replicates). 
 
 In the running bivariate lognormal example  with $\rho=0.7,$  the ET analysis has been explored for a range of choices of small values of the tolerance parameters $\tau,\epsilon$.  There is a high level of robustness with respect to these values.  Take  $\tau=0.005$ as in the accept/reject analysis, and, for example, 
 $\epsilon =0.001.$  Monte Carlo samples of $I = 1\times 10^6$ generate empirical distributions of resampled $\y^i$ values that are visually indistinguishable from the direct accept/reject results shown in  Figure~\ref{fig-bivarLNABC}. The resulting constrained medians of each element of $\y$ are equal to those from the traditional accept/reject analysis up to two decimal places (on the practically relevant scale of $0-20).$ 
 This, and multiple other examples, supports ET as a novel approach to decision-guided conditioning on almost-exact deterministic constraints.  

Concordance of ET with  probabilistic conditioning is also clear in questions about constrained values that are  extreme under $P(\y).$   One gauge of this in the ABC accept/reject analysis is the empirical estimate of acceptance rate $100 p_s\%.$ In the ET analysis, 
the effective \% sample size of the importance sampling weights, $\textrm{ESS}=100/\sum_i (I w_i^2),$ is comparable. 
 In the  example with  $\tau=0.005$ and $\epsilon = 0.001$ at the optimized values of $\gamma,$ the summaries are as follows. 
When $F=14.7$, the low value, the effective \% sample size of the importance sampling weights,  these two measures are each 0.56\% to two decimal places; 
when $F=24.15$, the high value,  they are each approximately 1.36\%. These very low values
are again indicative of the challenges of conditioning. Constraints that are unlikely under $P(\y)$ will 
generate low ESS and  showing the instability of the results.  In such contexts, taking $\epsilon$ very small and $m$ closer to $1$-- as required for theoretical reliance on the approach-- is increasingly fragile without access to very large samples from $P(\y).$    That said, it is very worthwhile to explore 
both probabilistic conditioning using ABC-style analysis and the ET approach together in a given context. The formal Bayesian decision analysis approach-- with its different goals and outputs--  generates additional results and insights, as is now further exemplified.

\subsubsection{Constraint Sensitivity Analysis} 

\begin{figure}[htpb!]
\centering
\vskip.5in
\includegraphics[width=0.475\textwidth]{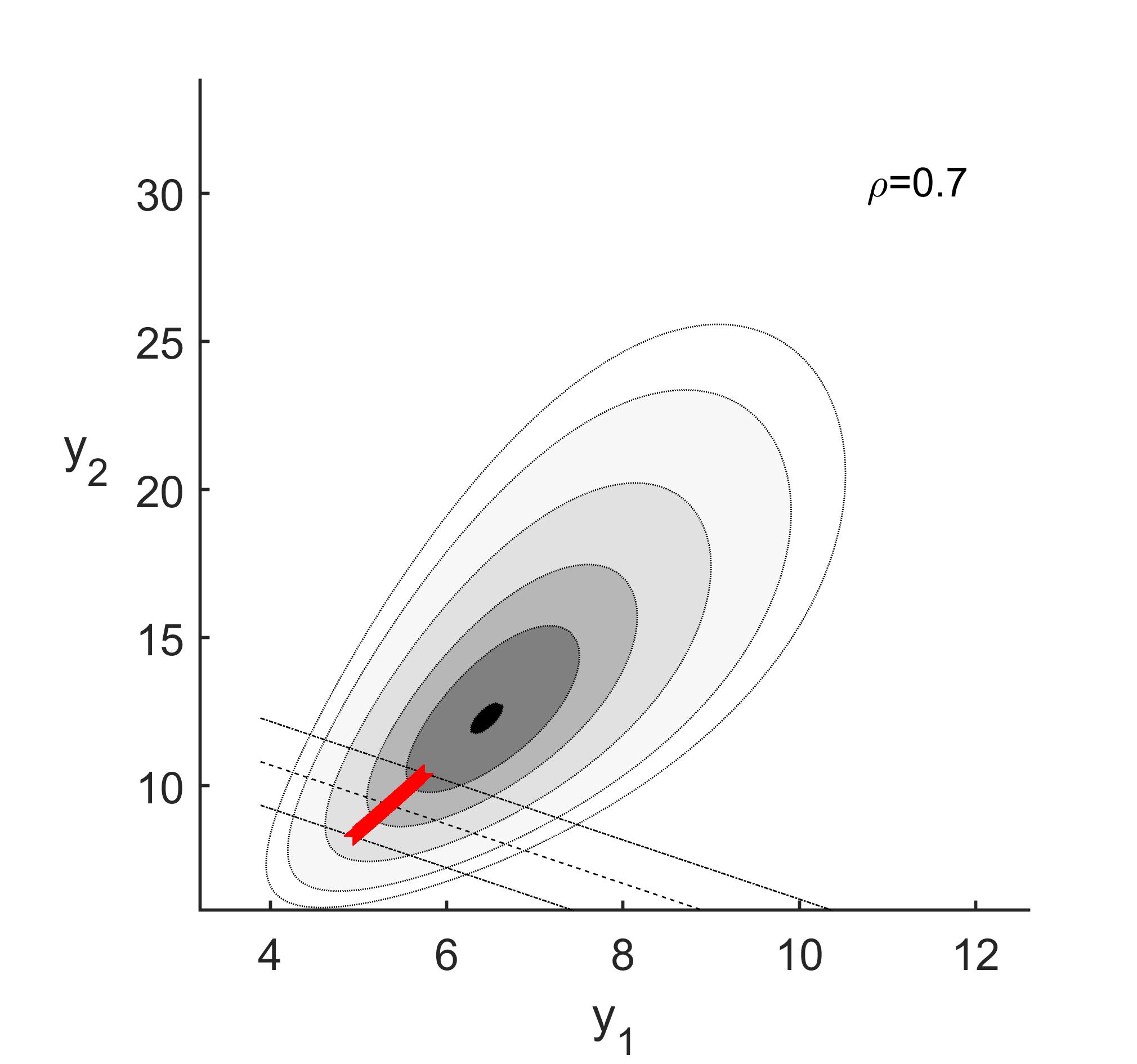}
\includegraphics[width=0.475\textwidth]{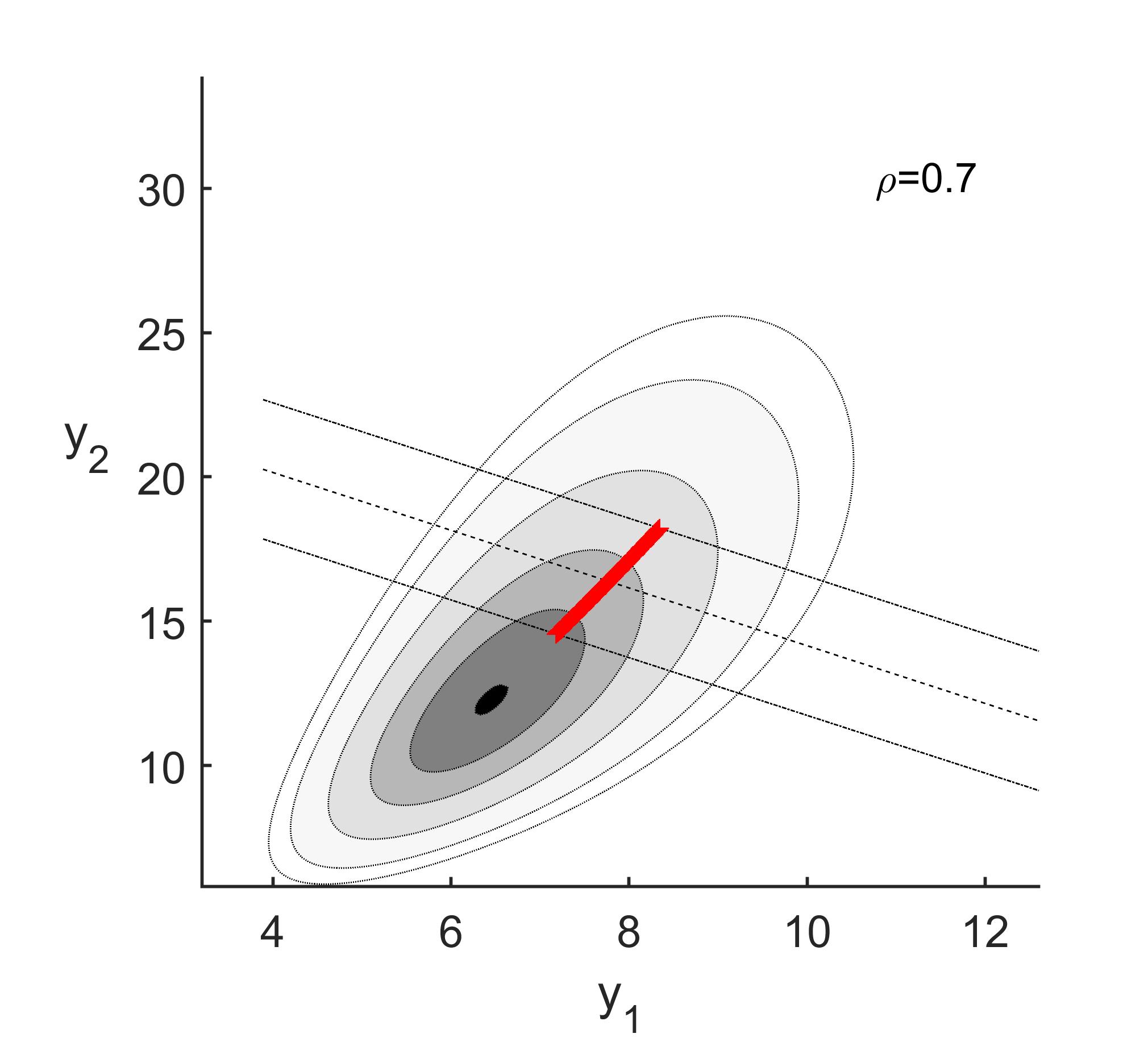}
\includegraphics[width=0.475\textwidth]{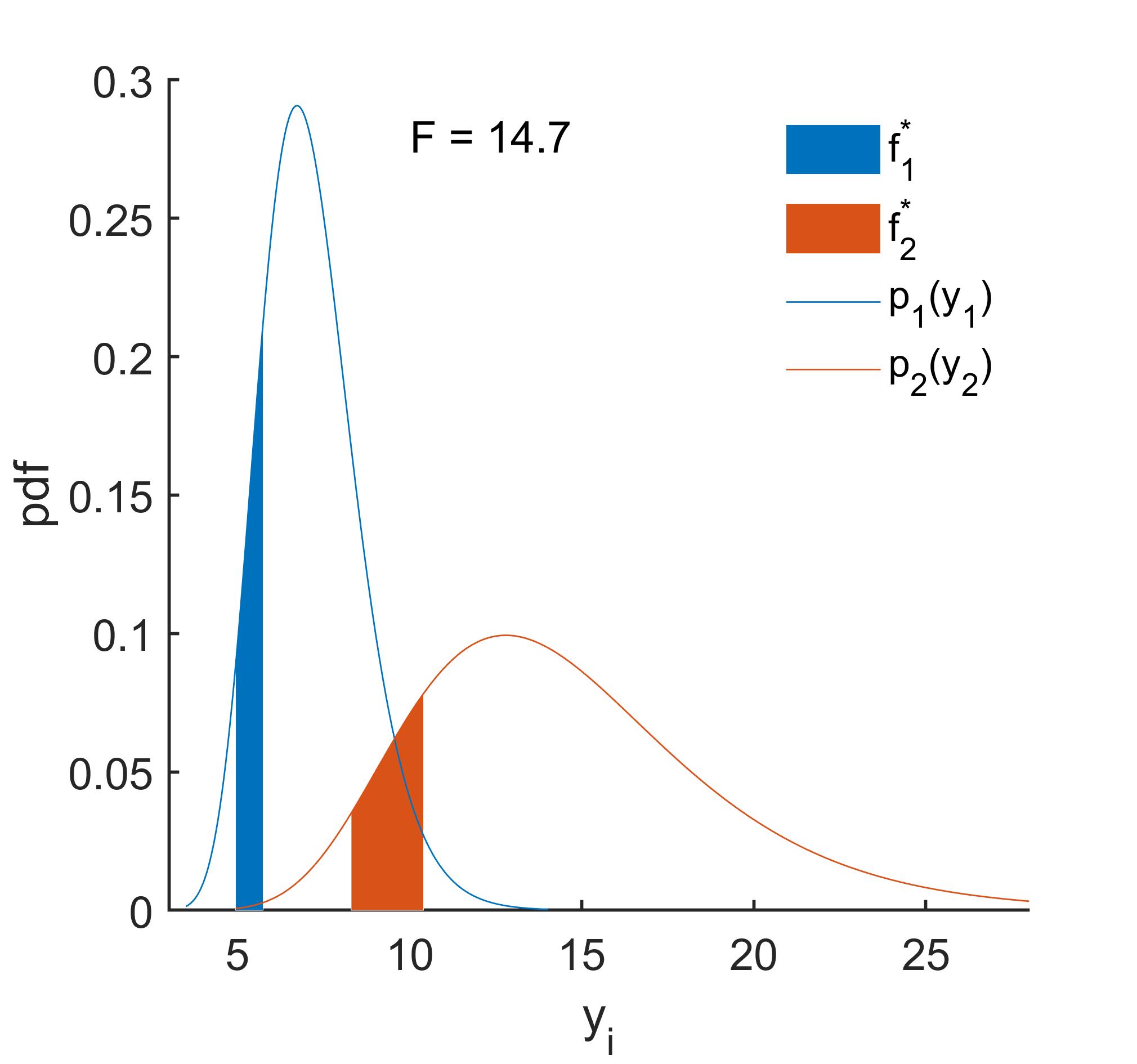}
\includegraphics[width=0.475\textwidth]{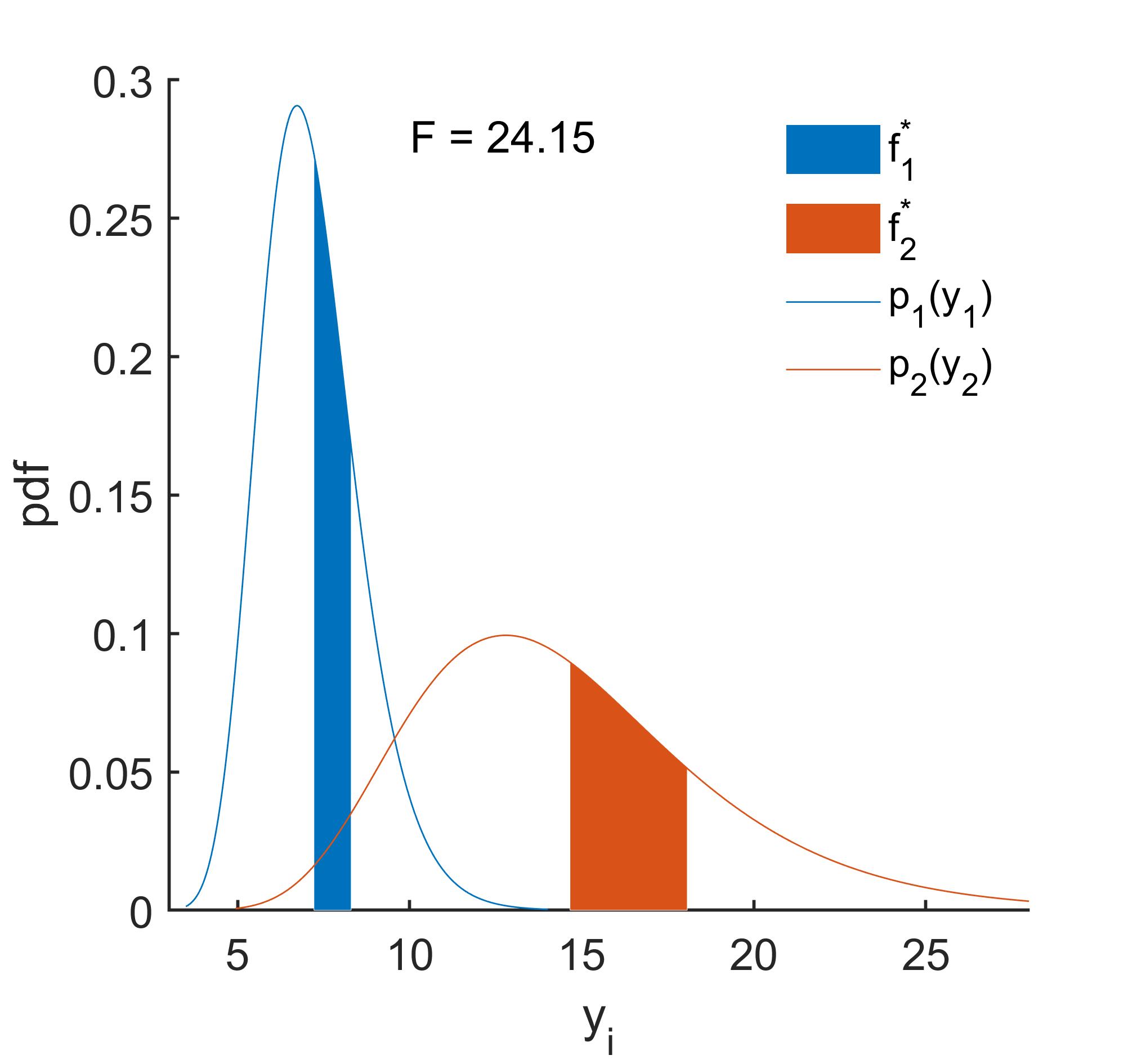}
\caption{Bivariate lognormal example with $\rho=0.7$, showing the joint p.d.f. and its marginals with constrained ranges of each $\optfi$ indicated. 
This is based on  the  sensitivity analysis under AD loss and  varying the 
constrained total in   $\{0.9F, 1.1F\}$-- i.e., within  $\pm 10\%$ of the nominal value $F$.  
Implications of the earlier used low ($F=14.7$, left figures) and high ($F=24.15$, right figures) values for the-- now nominal-- constrained total are shown. 
The upper frames show as dashed lines the nominal constraint with now lower and upper bounds also indicated. Superimposed 
is a scatter plot (+) of the $\optf$ as $F$ varies across its range. The lower frames show the marginal p.d.f.s with ranges of each $\optfi$ indicated by 
the shaded regions. \label{fig-bivarLNThresh} }
\end{figure}

Exploring  loss outcomes under perturbations of the chosen conditioning value $F$  defines a local sensitivity analysis: perturb a \lq\lq nominal'' constrained value $F$ and  reevaluate $\optf$ across perturbed values $F+\delta$ for some $\delta$ in a specified discrete range of values. In this example context with positive outcomes, perturbations of a chosen sum $F$ are best couched as percentage changes, i.e., taking $\delta = \pm\epsilon F$ for small $\epsilon$ on a discrete range of specified values. 
To illustrate this here, Figure~\ref{fig-bivarLNThresh} summarizes results in this bivariate lognormal example for two cases of the nominal $F,$  low and high values with respect to $p(Y)$, and taking $\epsilon=0.1.$  The resulting ranges of $\optf$ values as $F$ varies within 10\% of the chosen nominal values indicate quite tight ranges, in this example context. Note how the optimal point forecast values track the levels of the joint p.d.f. as $F$ varies,  with trajectories that appear to move very naturally along a \lq\lq ridge'' in the p.d.f. 
Also, while these regions of optimal point forecasts are of course very different conceptually to probability intervals under the purely probabilistic framework, note the concordance with the ABC-approximate conditional p.d.f.s in  Figure~\ref{fig-bivarLNABC}.  

The   NR algorithm is fast; running it for multiple values $F+\delta$   is computationally easy. A first-order approximation   is available to define a computational short-cut, perhaps at least for initial exploratory analysis.  Based on the NR update equation,  a perturbation of $F$ to $F+\delta$ for some small $\delta$ yields the update to $\lambda = \optlambda + \delta/\dot q(\optlambda).$ In the current example context this translates to   $\lambda = \optlambda + \epsilon F/\dot q(\optlambda)$ where $\epsilon$ takes value in a small range of \% changes.  This provides a trivial short-cut approximation to evaluating $\lambda$  over the range, and then computing the implied ranges
$\optf(\lambda)$ in the  sensitivity analysis.

Finally, in some applied settings it may be of interest to explore  sensitivity analyses with different ranges of values of $F$, such as defined by predictive intervals for $Y$ under some external model for the outcome total 
that is imposed on the predictive model $p(\y)$ in the spirit of information aggregation, or predictive synthesis~(\citealp{West1992c}; \citealp{West1992d};  \citealp{West1997}, section~16.3;  \citealp{McAlinnWest2017bpsJOE}).

\subsection{A $\mathbf{100-}$Dimensional Example} 

Some of the main motivating areas of application are in constrained forecasting in commercial or economic settings. 
 In commercial sales forecasting in large companies,  key example contexts including those of 
projecting point forecasts throughout hierarchies of sales or revenues, and of ensuring consistency of forecasts for sales at increasingly fine levels of disaggregation. Forecasts for \lq\lq high-level'' sales or revenue must be consistent with sets of forecasts at \lq\lq lower levels''~\citep[e.g.][section~16.3]{HarrisonGreen1973,West1997}. Here the general framework concerns conditioning sets of forecasts on information about totals and aggregates, all of which can be represented via sets of linear constraints on the uncertain outcomes being predicted.   Such questions arise commonly in \lq\lq what-if'' evaluation in policy decision contexts such as above and in other areas including  macro-economic forecasting over multiple time periods~\citep{McAlinnEtAl2019}. Related questions arise in ensuring compatibility  of forecasts at different resolutions in time, often with different forecast models generating predictions at different time scales~\citep[e.g.][]{FerreiraLeeBook2007,Molina2010,BerryWest2018TSM}.  

\begin{figure}[htpb!]
\centering
\includegraphics[width=0.475\textwidth]{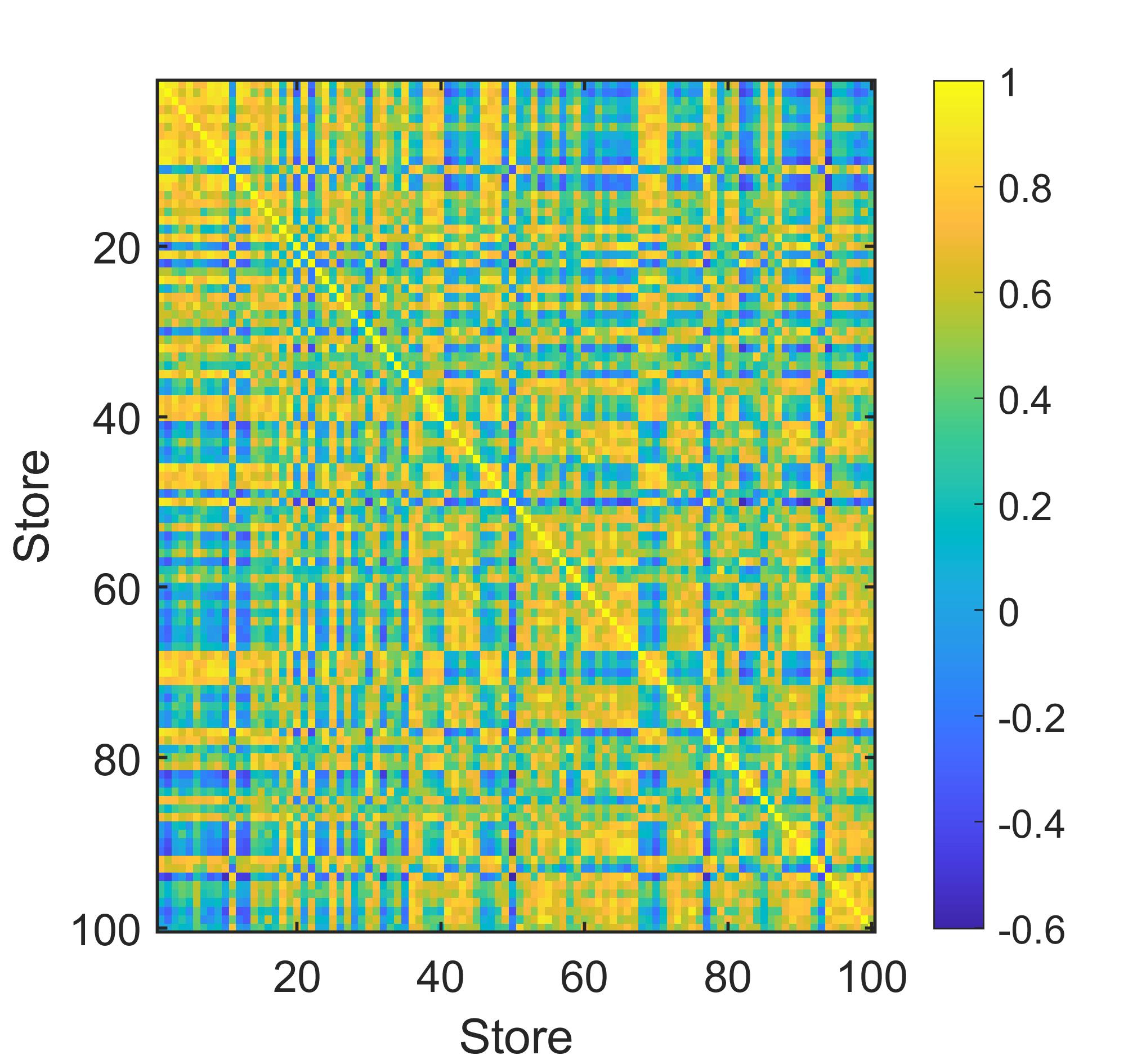}
\includegraphics[width=0.475\textwidth]{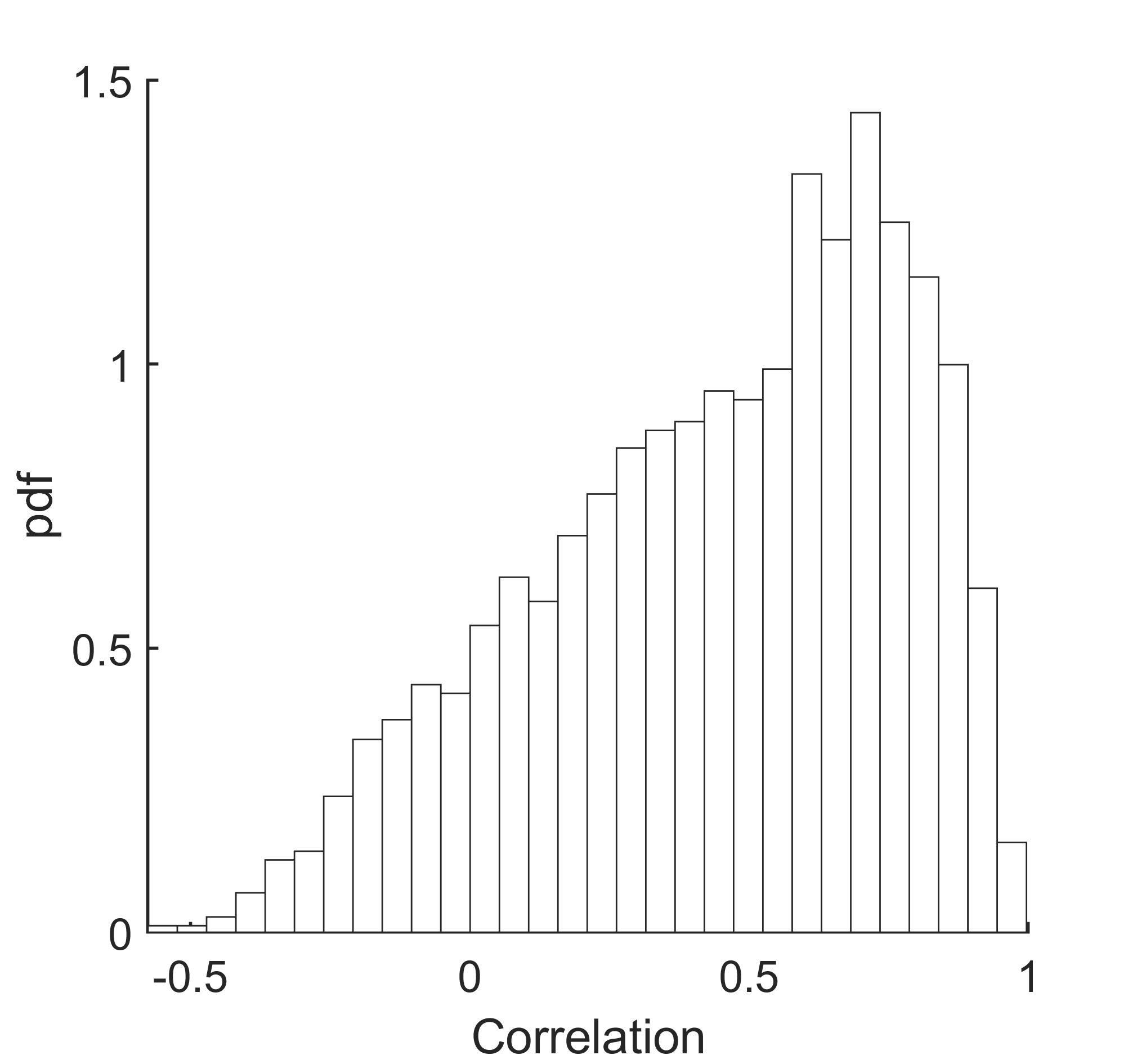}
\includegraphics[width=0.475\textwidth]{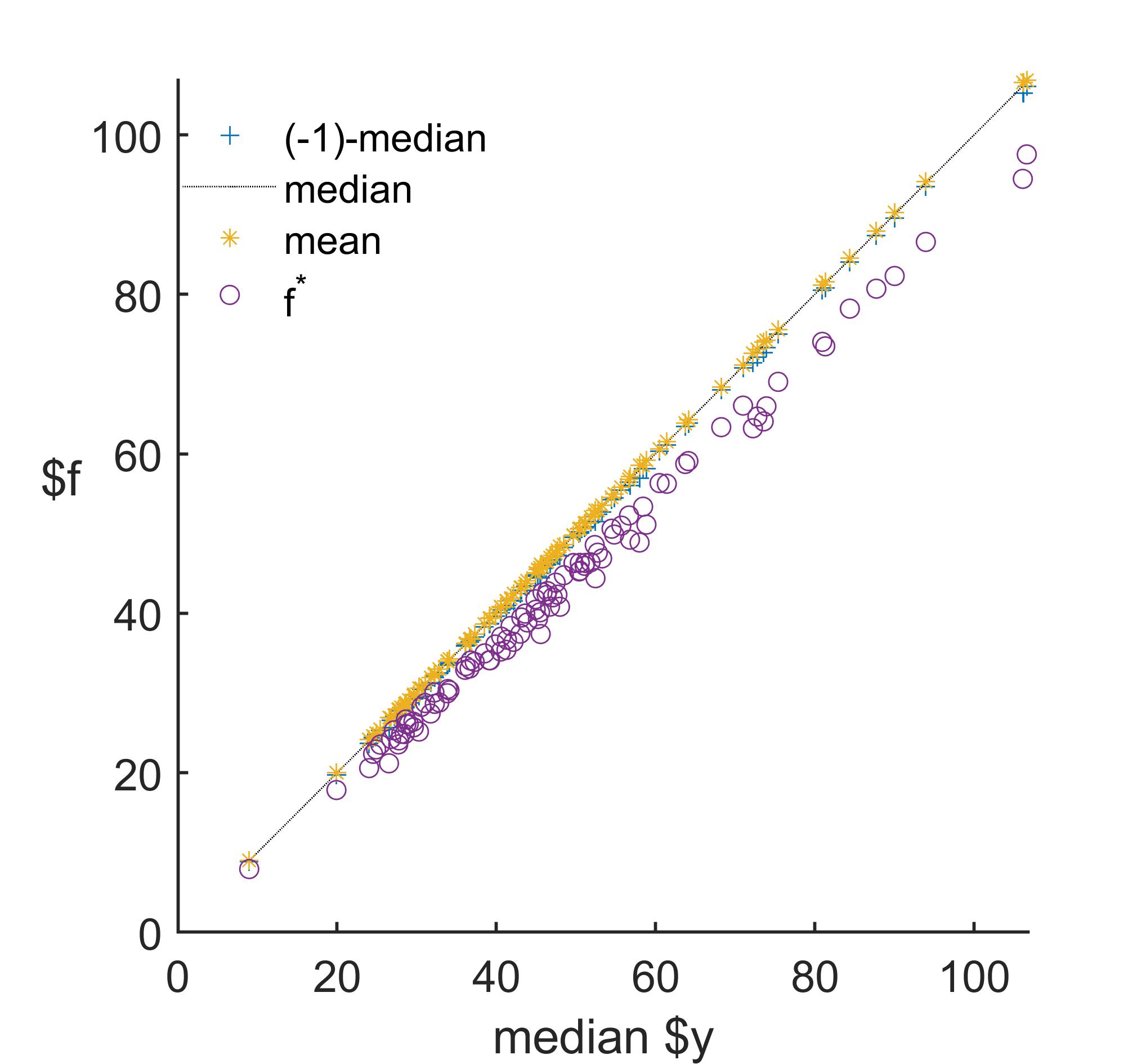}
\includegraphics[width=0.475\textwidth]{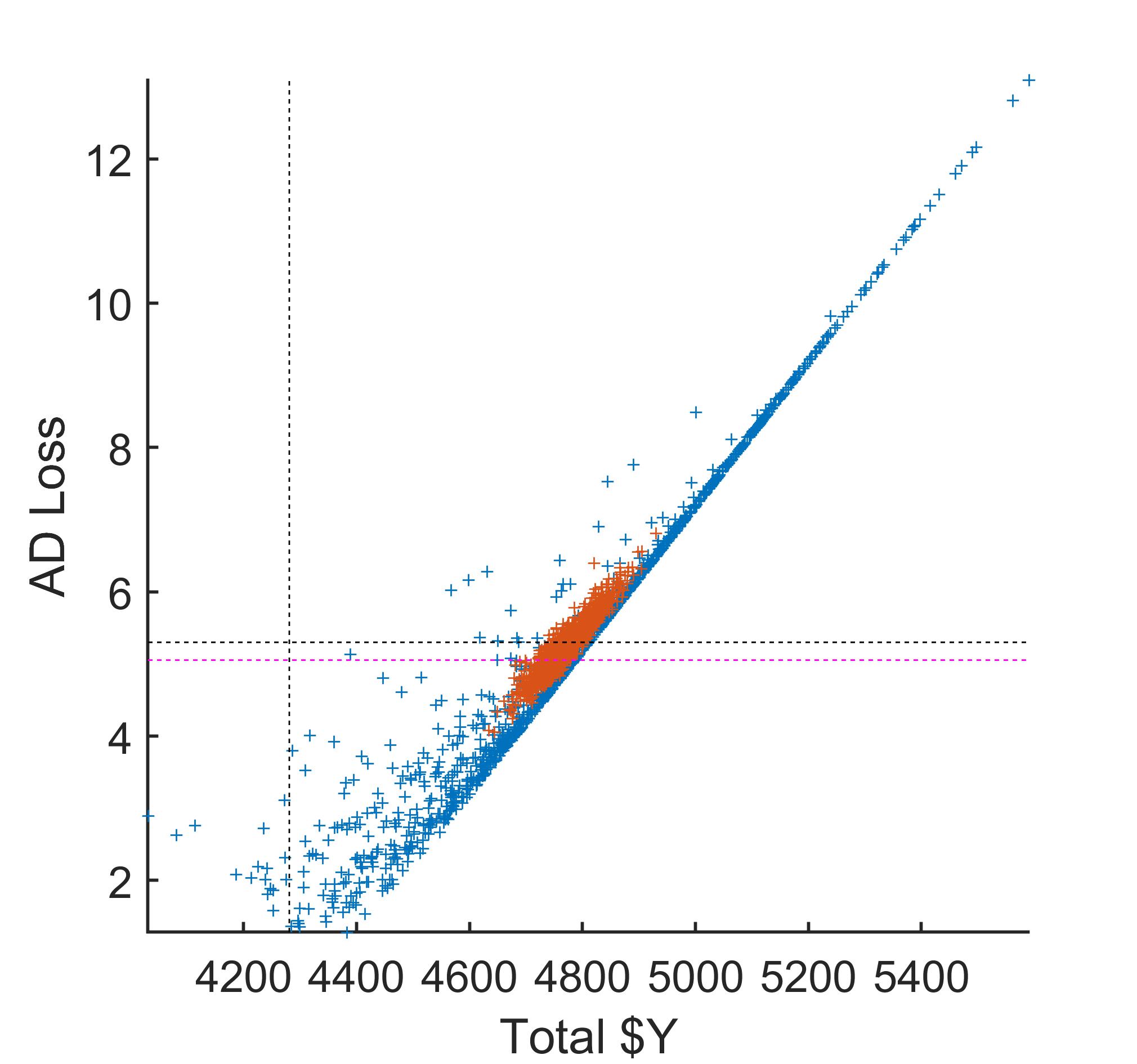}
\caption{Supermarket sector sales revenue example with $n=100$ and $\y\sim LN(\m,\V)$.   The top frames exhibit the correlations underlying $\V$ in heat-map and 
histogram forms. 
The lower left frame plots the marginal $(-1)-$medians, medians and means of the $y_i$ against their medians, and overlays a scatter plot of the AD-optimal $\optfi$ point 
forecasts in the case of a total constraint $F= 4{,}281.$ This value of $Y=F$ lies somewhat in the lower tail of $P(Y)$ as exhibited in the lower right frame that scatter plots 
a Monte Carlo sample from $P(Y,L(\y,\optf)/n)$ ({\em  blue +}).   The vertical dashed line marks the value of the constraint, $Y=F$; the horizontal dashed lines 
mark the value of the expected loss at the minimum, i.e., the optimized risk $R(\optf)$  (black dashed line) and the median of the loss distribution (red dashed line). 
Overlaid is a scatter plot (+) of a corresponding sample from a modified $P(\y)$ that has the same location and scale parameters but sets dependencies to zero, i.e., $\y \sim 
LN(\m,\V_0)$ where $\V_0$ has the same  diagonal elements as $\V$ but zero off-diagonal entries.
	       \label{fig-100stores} }
\end{figure}

An example in $n=100$ dimensions is summarized in Figure~\ref{fig-100stores}, linked to applied studies in supermarket sales modelling and forecasting. The data come from   $n=100$ stores with monthly revenue $\$y_i$ in the same consumer goods sector in each store, recorded each month over several years.  The data are scale transformed for confidentiality. 
The snapshot here concerns a forecast distribution $P(\y)$ where $\y$ represents the one-month ahead revenue vector at a chosen time point. 
Here $P(\y)$ is a $100-$dimensional lognormal $\y\sim LN(\m,\V)$ with $(\m,\V)$ set at values based on the historical record and model analysis. 
Conditioning on the sales total $Y$ is then of relevance from the above noted perspectives. A point forecast of $Y$ may be generated from an \lq\lq external'' source and the goal is to explore consistency with $P(\y)$ and/or to proceed to constrained forecasting of $\y.$ The external source may be an alternative model of aggregate sales, or one of a number of values generated as \lq\lq what-if'' or scenario values for consideration.  
A main point for illustration is to complement the above examples in highlighting the role of dependencies in a total-constrained decision analysis and the impact of the decision perspective. This is also a higher-dimensional example and it should be noted that the computational load in evaluation of decision analytic constrained forecasts remains almost trivial using the NR algorithm.

Figure~\ref{fig-100stores} displays summaries of   correlations in $\V$. There are   dependencies across stores, with 
 both negative and positive dependencies exhibited in the displays of  correlations   defined by $\V$.  
The decision analysis is summarized through evaluation of optimal point forecasts using absolute deviation (AD) loss; revenue outcomes are all on the \$ scale so are directly 
comparable and APE loss is less relevant, while the context is such that results under AD, APE and SE are in any case similar.  
Marginal point forecasts for each $y_i$ are closely similar across stores $i,$ as illustrated in the figure.  Using AD loss and constraining to the total $F=\one'\f$ 
that is fixed at a value somewhat (though not extreme) in the lower tail of $p(Y)$ shows optimal constrained forecasts that are clearly downward adjustments to the 
marginal values.   The figure then displays Monte Carlo samples from $P(Y,L(\y,\optf)/n)$ as in the earlier examples of Section~\ref{sec:bivarLNexamples}. 
That is, a scatter plot of samples from the joint distribution of the total revenue over stores together with the realized loss {\em per store} under the AD loss function. The imposed conditional total value here is 
$F= 4{,}281,$ lying in the tail-- though not really substantially extreme-- of the forecast distribution $P(Y)$ that is close to symmetric with median $4{,}775.$

The distribution of loss at the AD-optimal value $\optf$ is spread over $2-12$ on the scale defined in this analysis, while the mean and median of the loss distribution are around $5.1-5.2$. There is appreciable probability of loss values much less than this, as well as   reasonable chances of  higher losses up to the $9-12$ range (akin to \lq\lq value at risk'').   These are key and potentially critical presentations of realistic outcomes from the decision analysis, and simply proceeding on the basis of traditional \lq\lq act on the optimal decision'' does not recognize  these potentially important practical considerations.   

A further point   speaks to the impact of dependencies in $P(\y)$  on the implied loss distributions. 
The figure highlights this in contrasting summaries of the predictive distribution $P(Y,L(\y,\optf)/n)$  under the dependent model with a  model in which $\V$ is replaced by a diagonal matrix $\V_0$ having the same diagonal elements. This is not a strange choice for  comparison;  applied analyses of such problems will often  analyze data  independently across stores, so this is a relevant benchmark.  The resulting $P(Y)$ and hence the joint $P(Y,L(\y,\optf)/n)$ are very concentrated relative to the original model analysis. In the dependent model,  there are ranges of negative and positive dependencies among the $y_i$, but the preponderance and magnitudes of positive values lead to overall increased uncertainty about $Y$ and hence about the potential loss outcomes.   This is  typical in such commercial applications, where dependencies often arise through common factors such a seasonality and management policies that are comparable across stores~\citep{BerryWest2018DCMM,BerryWest2018TSM}.    The comparison analysis that fixes all correlations in $\V_0$ to zero defines, in  contrast, a rather concentrated predictive distribution for $Y$ and hence losses.  While the optimized expected losses are the same under the two models, the independence model massively understates the levels of realistic uncertainty in the outcome total $Y$ and hence in the loss distribution; this leads to the potential to generate substantial over-confidence in the selection of the optimal constrained point forecasts.    Other comparisons could be made, but this practically-grounded example serves to again highlight the main point of examining loss distributions along with optimal point forecasts.

\section{Summary Comments \label{sec:conclusion}} 
 
Integrating decision analysis into problems of constrained forecasting is complementary to  the traditional  inferential view, and examples here illustrate the benefits of ensuring attention to the decision theoretic \lq\lq Yang'' of Bayesian analysis as well as to the inferential \lq\lq Yin''. This view is, of course, not specific to the contexts of Bayesian constrained forecasting, but decision perspectives are  under-regarded and under-represented in applied forecasting as in other areas~\citep[e.g.][section 2.3]{Lindley1992ValenciaIV,LavineLindonWest2019avs,West2020Akaike}.    The potential for methodological advance as well as more comprehensive analysis is highlighted in the motivating context of constrained forecasting. Examples with features faithful to applied settings demonstrate the relevance of the broader view of decision analysis that expands from a main focus on optimal decisions to always explore implied loss distributions. Constrained forecasting examples where multivariate dependencies  can have major impact on ranges of likely losses highlight the practical import.     It is also noted that, in many common, practical contexts, expected loss functions are undefined while distributions of losses are   valid and accessible. Key examples  involve log-T distributions that routinely arise in linear models of many kinds, including dynamic linear models (DLMs) in time series forecasting. Here blind adoption of optimal point forecasts based on minimizing expected losses is unfounded implied predictive distributions of losses are perfectly well defined.

 Methodologically,  connections of decision analysis and inferential approaches are nicely enhanced by new ideas and methodology exploiting 
entropic tilting. As an approach to understanding compatibility of constraints with a forecast distribution, and to minimally modifying the distribution to approximately accord with  deterministic constraints, ET is a fully Bayesian decision analysis framework that can be exploited to extend methodology in practical ways. Its connections to  Monte Carlo methods for probabilistic conditioning nicely highlight its use and complementarities relative to constrained point forecasting alone.   Technically, use of more elaborate, non-linear constraints will raise challenges of optimization. While many, general algorithms exists for numerical optimization problems, linkages with Bayesian emulation approaches such as have been used in portfolio optimization subject to various constraints~\citep[e.g.][] 
{IrieWest2018portfoliosBA} may provide relevant and novel opportunities for progress.

Detailed application in live forecasting require extensions to contexts of multiple constraints such as arise in hierarchies. Extension of the basic decision problem of \eqn{constrainedoptimization} to  a set of constraints is theoretically immediate. That is, 
the optimization is generalized to condition on  $k$ constraints $\A'\f = \F$ where $\A$ is a given  $n\times k$ matrix of full rank $k<n$ and $\F$ a given constraint $k-$vector.   Developments with intersecting sets of subtotal constraints-- in which case $\A$ is a matrix with zero/one entries-- are one main class of interest.  Evaluation of optimal constrained forecasts  based on  multivariate  Newton-Raphson  is immediate.

Problems in which the loss functions are not additive in outcomes $i=\seq 1n$ are also of interest.  In commercial forecasting with positive, or non-negative, outcomes (such as with consumer sales of items or batches of items, revenues in multiple sectors or markets) it can be relevant to consider loss functions that involve cross-talk between outcomes.  Sales of one product may be inversely related to those of another due to substitution effects, and overall sales across categories might be of main interest; similar comments apply to revenue forecasting over multiple sectors.  Customized losses $L(\y,\f)$ that are not additive will be of interest. 

Important extensions concern problems in which the predictive distribution $P(\y)$ is itself impacted by the future actions based on chosen optimal point forecasts. 
As one example,   consider supermarket sales where $y_i$ is the sales outcome, and $\optfi$ defines the store manager's decision to stock $\optfi$ (or, perhaps, $\optfi+$ a few more) items of a specific consumer item.  Then, necessarily $y_i\le \optfi$ since no more than that are available for sale in the next time period.   Extension of this applies to cases of other constraints on $\y$, such as are relevant in macro-economic forecasting where one or more of the $y_i$ are putatively controllable as policy instruments.  A macro-economic forecasting model is then applied across a range of \lq\lq what-if?'' values of one variable, corresponding to one (linear) constraint on the outcome vector rather than a total constraint. Such conditional forecasting across multiple time periods  lies at the heart of applied Bayesian forecasting in macro-economics~\citep[e.g.][]{DelNegro2008,NakajimaWest2013JBES,McAlinnEtAl2019}.   This argues for extensions in which $P(\y)$ is modified to 
$P(\y|\F)$ in the constrained optimization setting. 
This is not a new concept~\citep[e.g.][]{HarrisonSmithValenciaI} but is certainly under-regarded in both forecasting and Bayesian analysis literatures. It is an extension of significant potential practical importance. 

 This paper also relates to  the broader area of  forecast combination and reconciliation, and notably  to the Bayesian literature  on forecast calibration and combining forecasts from multiple, potentially related sources.  This line of literature~(e.g., \citealp{West1992c}; \citealp{West1992d};  \citealp{West1997}, section~16.3;  \citealp{McAlinnWest2017bpsJOE}) intersects intimately with the apparently   more specific goals of constrained forecasting, but as detailed in examples in~\citet[][section~16.3]{West1997},   forms part of a broader context. Decision analytic perspective  of this paper can be 
expected to be applicable in these broader settings,.

\bibliographystyle{chicago}
{}

\newpage\setcounter{page}{0}\thispagestyle{empty}

\begin{center} 
{\bf \Large Perspectives on Constrained Forecasting}

\medskip
{\bf \large Supplementary Material: \\  \smallskip
 Further Technical Details and Additional Examples
} 

\bigskip
		{\em {\large{\bf Mike West\footnote{Department of Statistical Science, Duke University, Durham NC 27708, U.S.A.,~mike.west@duke.edu}}}}

\today

\end{center}

\subsection*{A.  Derivations for AD, APE and ZAPE Losses} 
 
 Additional details of risk functions under the various loss forms of Section~\ref{sec:decisionanalysisexamples}  are summarized here. The details 
 assume continuous forecast distributions  $P(\y)$ so that routine calculus defines the optimization analysis.  The results noted in the paper are parallel in cases of discrete (or indeed mixed discrete and continuous) distributions, with finite differences replacing differentiation in the derivations. 

Key ingredients of analysis under loss functions involving absolute forecast errors are the following results.  For each $i=\seq 1n,$  take any continuous 
distribution $H_i(y_i)$ with p.d.f. by $h_i(y_i).$  Define 
$$\rho_i(f_i) = \int_{-\infty}^\infty |y_i-f_i| dH_i(y_i),$$ assumed to be finite for all $f_i.$  Then  
\begin{align*} \rho_i(f_i)   &=  \int_{-\infty}^{f_i}  (f_i-y_i) dH_i(y_i) + \int_{f_i}^\infty  (y_i-f_i) dH_i(y_i)\\
		     &=  f_i H_i(f_i) - \int_{-\infty}^{f_i}  y_i dH_i(y_i) + \int_{f_i}^\infty  y_i dH_i(y_i) - f_i \{ 1-H_i(f_i) \} \\
		     &=  2f_i H_i(f_i) - f_i - \int_{-\infty}^{f_i}  y_i dH_i(y_i) + \int_{f_i}^\infty  y_i dH_i(y_i). \\
\end{align*} 
Differentiating with respect to $f_i$ yields derivative 
$$\dot\rho_i(f_i) = 2H_i(f_i)-1.$$ 
This feeds into the constrained optimization results for:   (i) AD loss, with $H_i(\cdot)=P_i(\cdot); $  (ii) for APE loss, 
with $H_i(\cdot) = G_i(\cdot)$ defined via p.d.f. $g_i(y_i) \propto p_i(y_i)/y_i;$ and (iii) for ZAPE loss,  with 
$H_i(\cdot) = G_i(\cdot)$ defined via p.d.f. $g_i(y_i) \propto p_i^+(y_i)/y_i.$  Specifics are noted. 

\subsubsection*{AD Loss} 
Under AD loss, minimization of the constrained risk of~\eqn{constrainedriskfunction} with additive risk function of~\eqn{additiveriskfunction} is achieved by solving $\dot R_i(f_i) = \lambda$ for $i=\seq 1n$ subject to $F=\one'\f,$  where $R_i(f_i)=\rho_i(f_i)/c_i$ with 
$H_i(\cdot)=P_i(\cdot).$  This reduces to $2 P_i(f_i)-1 = \lambda c_i$ so 
$\optfi(\lambda) = P_i^-((1+\lambda c_i)/2).$ 

\subsubsection*{APE Loss} 
Under APE loss, minimization of the constrained risk of~\eqn{constrainedriskfunction} with additive risk function of~\eqn{additiveriskfunction} is achieved by solving $\dot R_i(f_i) = \lambda$ for $i=\seq 1n$ subject to $F=\one'\f,$  where $R_i(f_i)=\rho_i(f_i)/(c_ik_i)$ with 
$H_i(\cdot)=G_i(\cdot)$   defined via p.d.f. $g_i(y_i) = k_i p_i(y_i)/y_i.$  
This reduces to $2 G_i(f_i)-1 = \lambda c_ik_i$ so 
$\optfi(\lambda) = G_i^-((1+\lambda c_ik_i)/2).$

\subsubsection*{ZAPE Loss} 
Under ZAPE loss, minimization of the constrained risk of~\eqn{constrainedriskfunction} with additive risk function of~\eqn{additiveriskfunction} is achieved by solving $\dot R_i(f_i) = \lambda$ for $i=\seq 1n$ subject to $F=\one'\f,$  where 
$R_i(f_i)= \pi_{i0}f_i/c_i +  (1-\pi_{i0})\rho_i(f_i)/(c_ik_i)$ with $\rho_i(\cdot)$ and $k_i$ now based on  
$H_i(\cdot)=G_i(\cdot)$   redefined via p.d.f. $g_i(y_i) = k_i p_i^+(y_i)/y_i.$  
This reduces to $2 G_i(f_i)-1 = k_i(\lambda c_i-\pi_{i0})/(1-\pi_{i0})$ assuming the right-hand side expression lies in $[0,1).$ 
It follows that 
$$
\optfi(\lambda) = \begin{cases} 0, & \textrm{if } u_i(\lambda)\le 0, \\  G_i^-(u_i(\lambda)), & \textrm{if } u_i(\lambda)> 0,\end{cases}
\quad\textrm{where}\quad  
u_i(\lambda) = \frac 12 \left\{1+ k_i \frac{(\lambda c_i-\pi_{i0})}{(1-\pi_{i0})}\right\}.
$$

\subsection*{B.  Log-T  Distributions} 

\begin{figure}[t!]
\centering
	\includegraphics[width=0.475\textwidth]{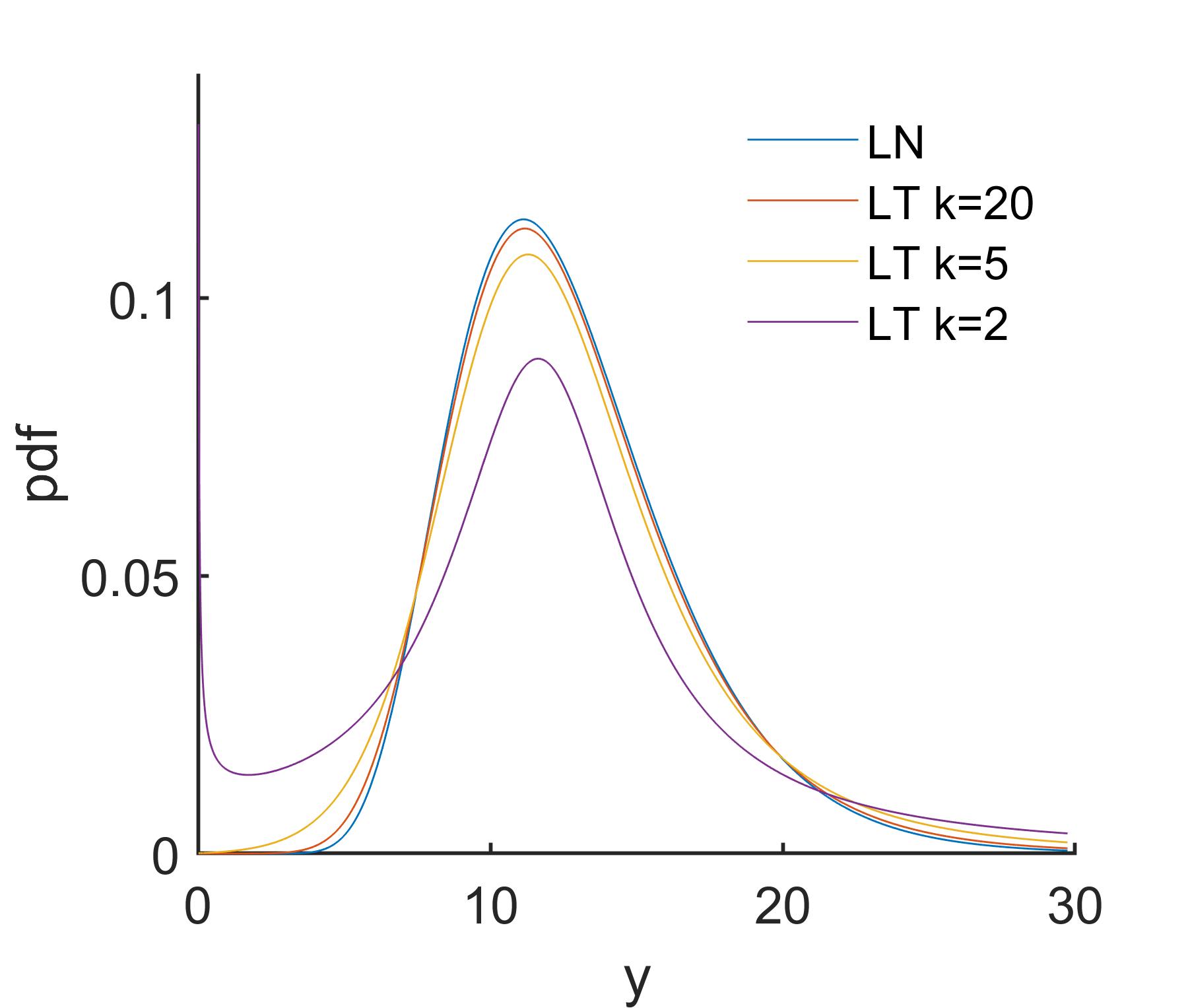}       \caption{Probability density functions of the $LN(m,v)$ and $T_k(m,v)$ distributions
	   with $m=2.5, v=0.09,$ and for $k=20,5,2.$ \label{fig:LTpdfs}}
\end{figure}

If $y=\log(x)$ and $x\sim T_k(m,v)$ then $y\sim LT_k(m,v),$  a  heavy-tailed log-T distribution with p.d.f. 
$$ p(y) \propto y^{-1} \{ k+ (\log(y)-m)^2/v \}^{-(k+1)/2}, \qquad y>0.$$
The p.d.f. decays like an inverse power of $\log(y)$ for $y\to \infty,$ and also-- perhaps initially surprisingly-- has  pole at zero.  The very heavy left tail of $p(x)$ 
transforms to mass compressed just above zero under $p(y),$ leading to a mode at zero with infinite p.d.f. 
This makes $p(y)$ bimodal in many cases (so long as $m$ is large enough). Otherwise, the p.d.f. has a shape that appears similar to the unimodal lognormal, and for larger degrees of freedom $k$ the pole at zero is hard to see in graphs, but is always there.   See Figure~\ref{fig:LTpdfs}.  In terms of location and other summaries, the median is $\exp(m)$ of course, and all other quantiles transform similarly directly from those of the T distribution. 
As a result of the heavy-tailedness, no moments exist.  If $p(y)$ has a moment generating function (m.g.f.) it would be 
$E[\exp(ty)]$ as a function of real values $t.$    For $t>0$, the implied integrand blows up exponentially as $y\to \infty$, since the exponential dominates the 
inverse powers of $\log(y). $  For $t<0$ the integrand also blows up exponentially as $y\to 0. $   Hence the m.g.f. does not exist, the distribution having no positive or negative moments.  Random samples from $p(y)$  will, of course, have finite realized values of sample moments, but their sampling distributions   have infinite moments and taking $k$ smaller quickly shows the sample summaries increasing-- as they are theoretically guaranteed to do-- to infinity. 
\FloatBarrier

A serious practical implication is that, under commonly arising predictive log-T distributions, expected losses using many of the standard families of loss functions are undefined/do not exist.  For example, means and variances are infinite so that quadratic loss is irrelevant. Similarly, AD, APE and ZAPE losses have no finite expectations. Of course, quantiles functions are well-defined so that the derivations arising in use of these loss functions can be applied to derive point forecasts, albeit without the explicit optimality property that they have when expected loss functions are finite.      Two other practical points are that: (i) simply truncating a log-T (or other) distribution to a bounded ranges away from zero to some finite upper bound leads, of course to finite expected losses; (ii) an alternative is to modify the loss functions so that they are themselves bounded. 


\subsection*{C.   Additional Examples \label{sec:moreexamples}}

\begin{figure}[p!]
\centering
\includegraphics[width=0.375\textwidth]{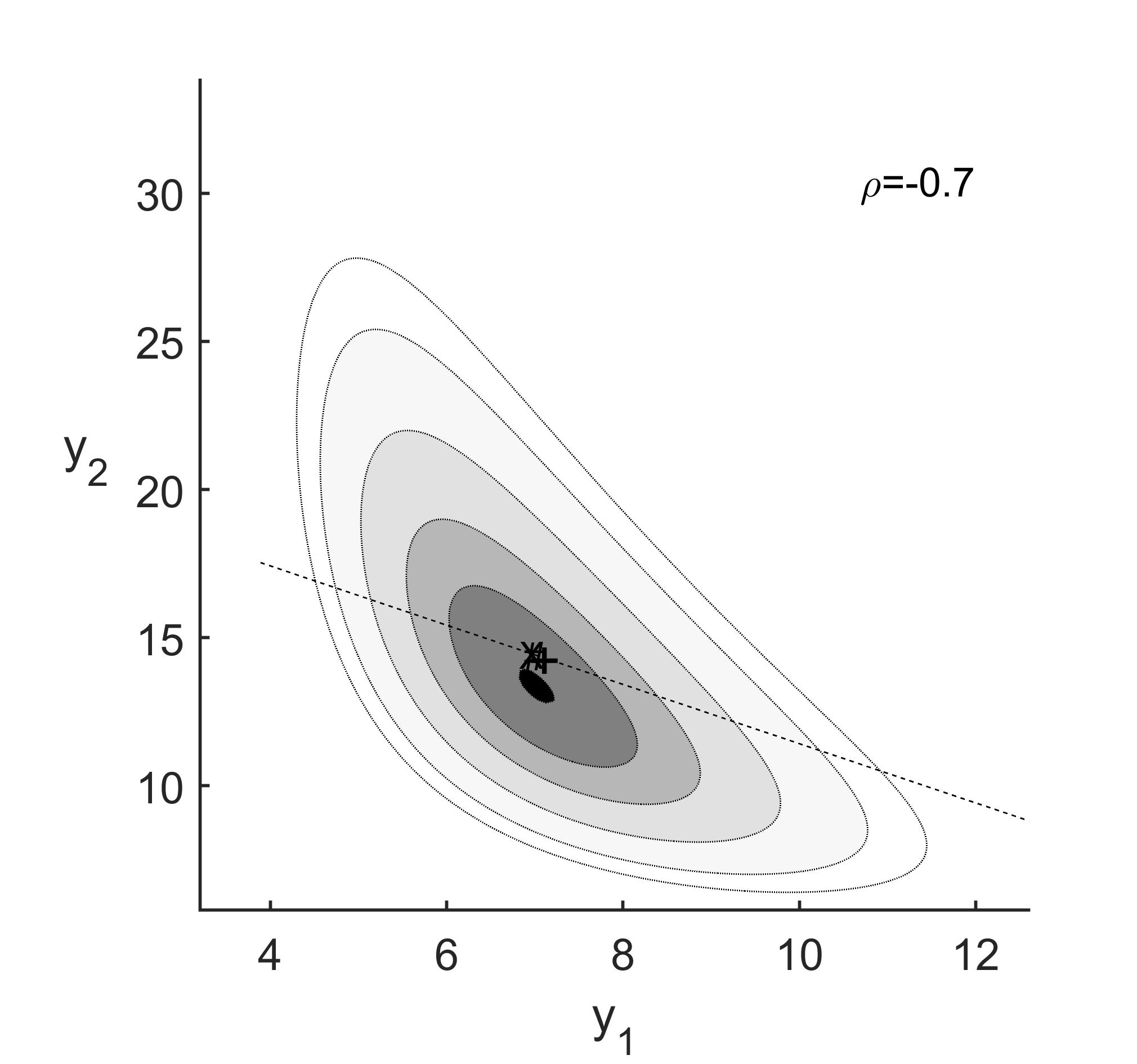}
\includegraphics[width=0.375\textwidth]{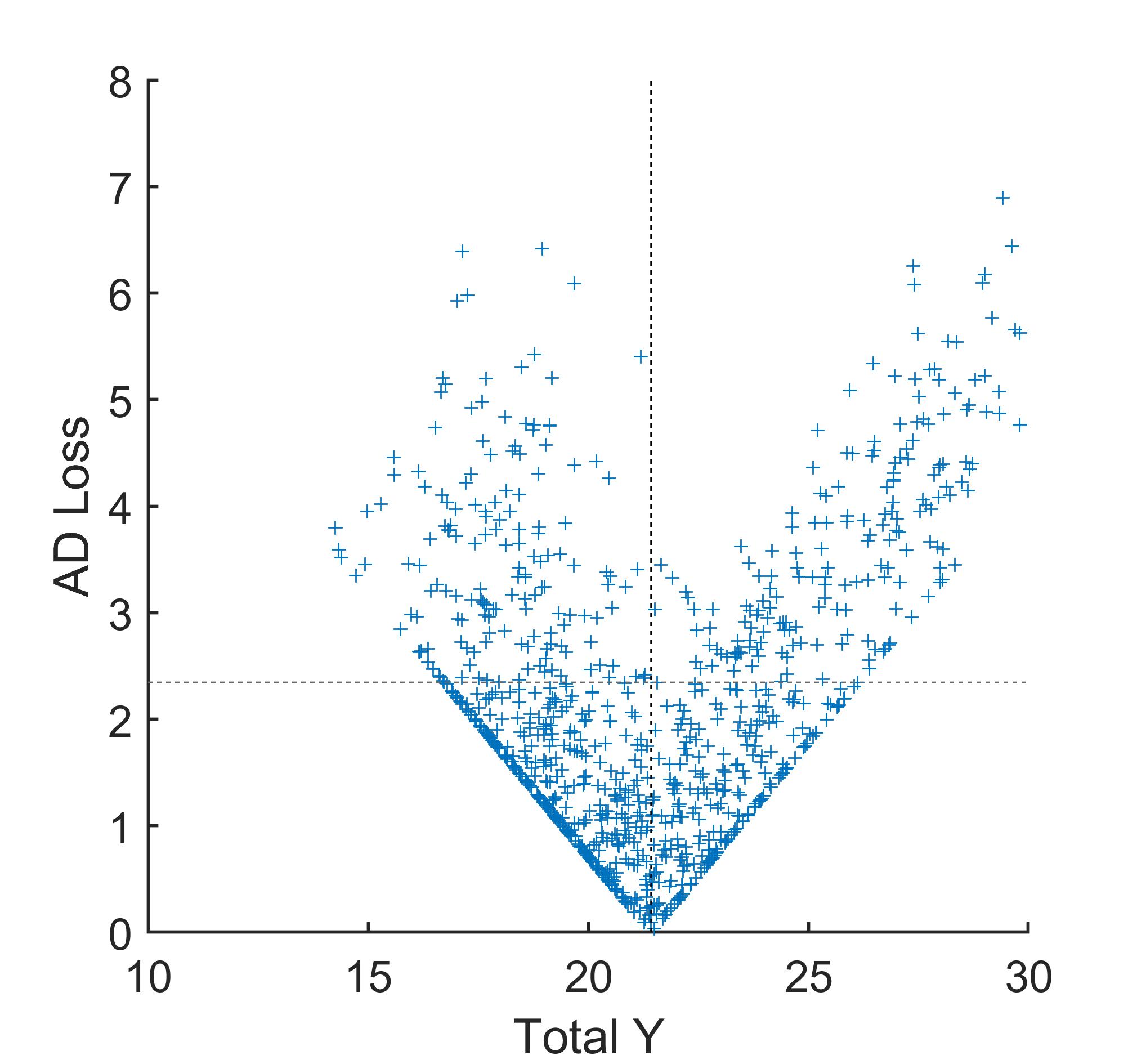}
\includegraphics[width=0.375\textwidth]{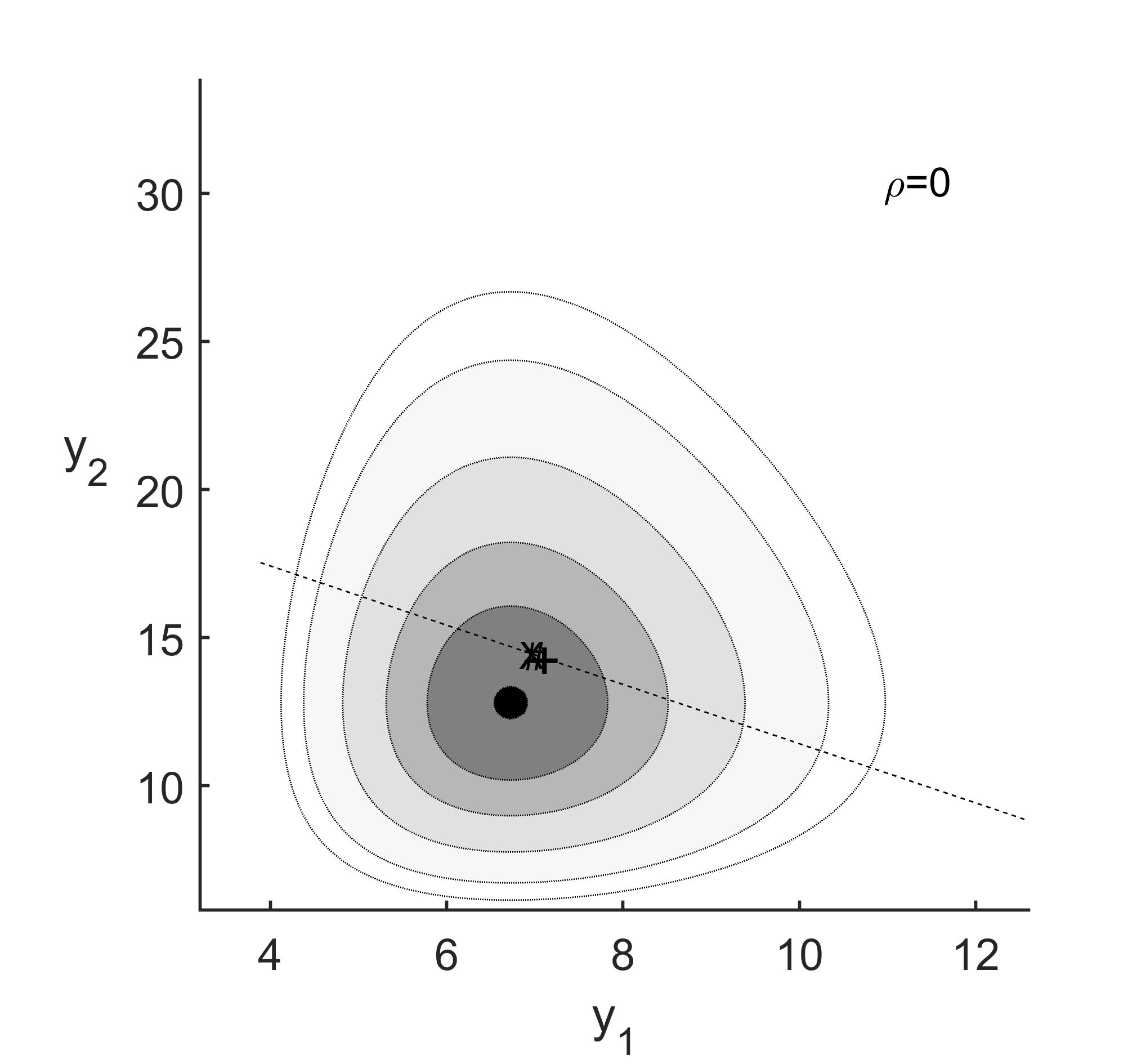}
\includegraphics[width=0.375\textwidth]{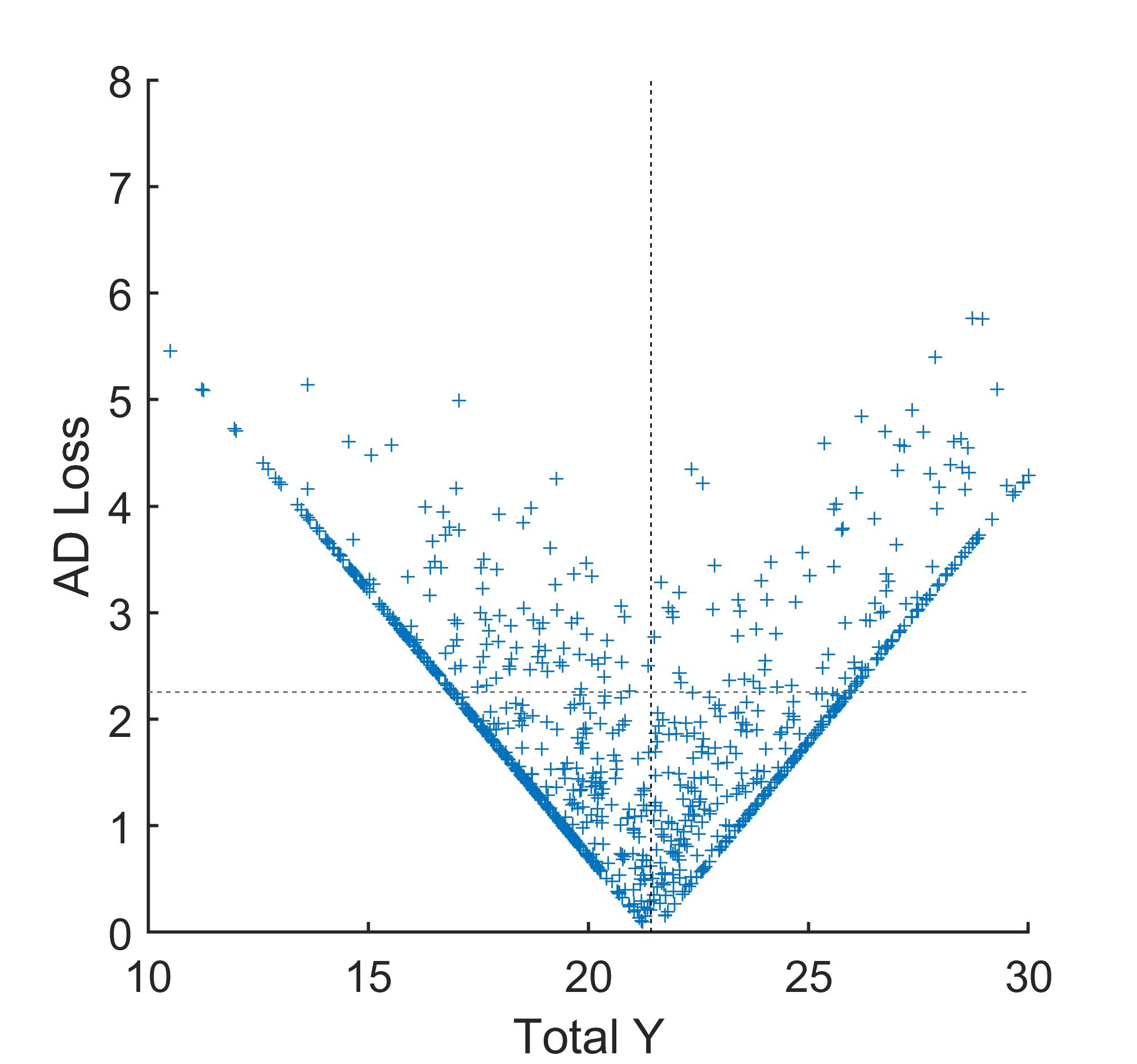}
\includegraphics[width=0.375\textwidth]{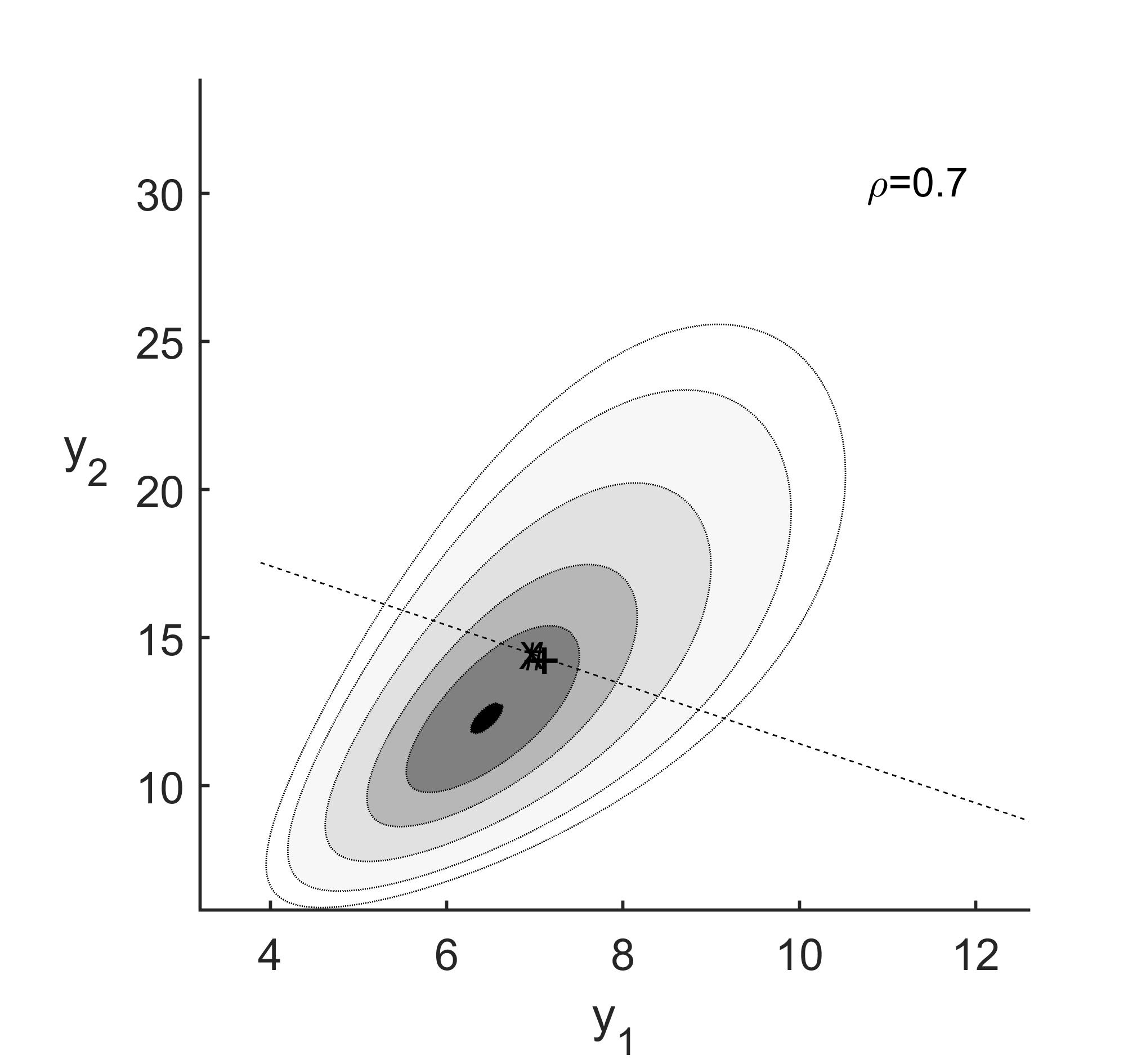}
\includegraphics[width=0.375\textwidth]{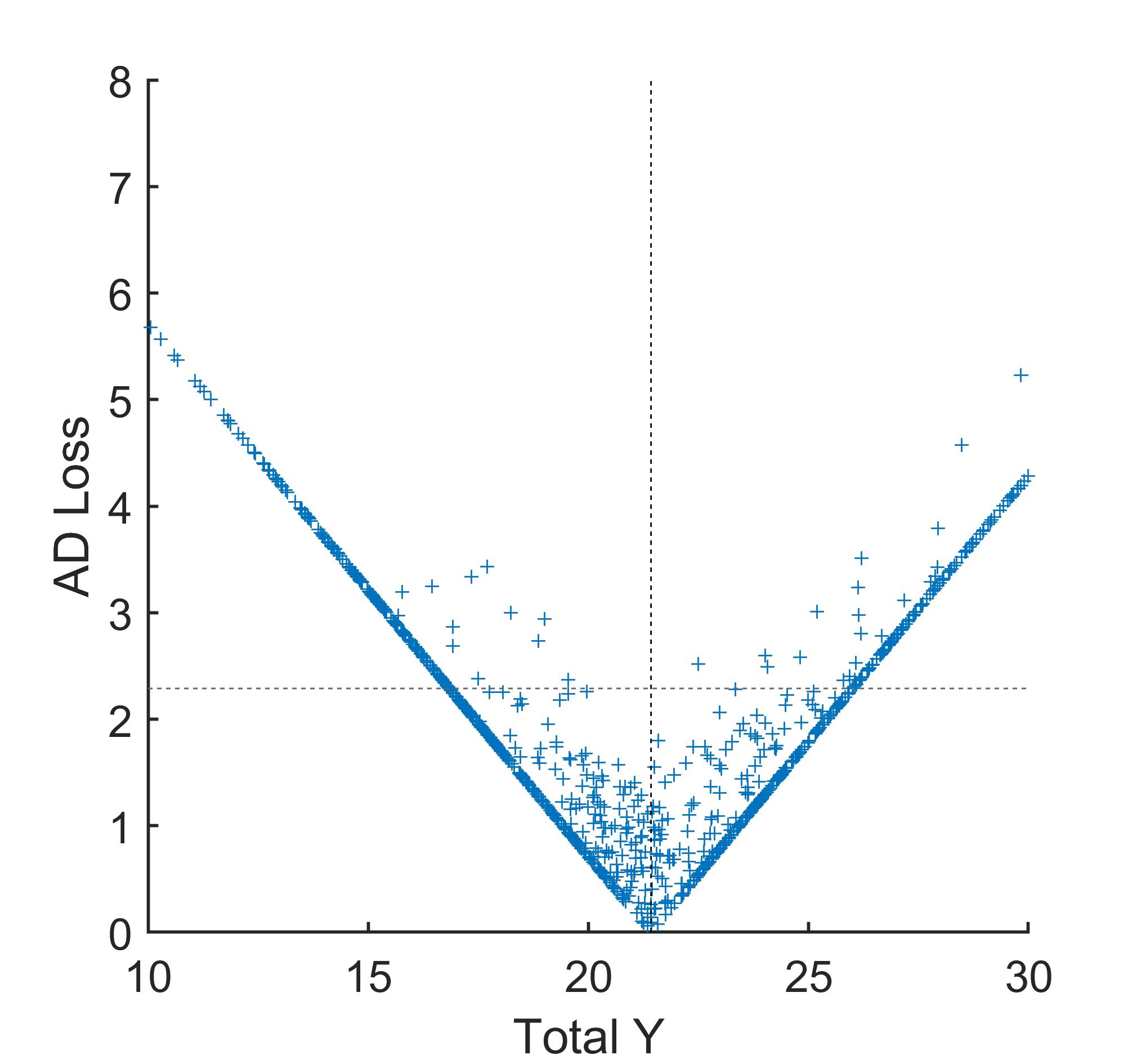}

\caption{Bivariate lognormal example as in Figure~\ref{fig-1bivarLNcontoursandlossdstn} of the text, but now with the value $F=21.4$ that-- lying close to the center of $P(Y)$-- is highly concordant with the predictive distribution $p(\y).$   {\em Left column:}  Contours of $p(\y)$ for three different levels of dependence $\rho\in \{-0.7,0,0.7\}$ and with the constraint $\one'\y=F$ indicated as the dashed line.  The symbols indicate the optimal point forecast vector $\optf$ under 
AD loss (+), APE loss (\#) and SE loss (X).   {\em Right column:}  Scatter plots of the corresponding   a Monte Carlo sample from $P(Y,L(\y,\optf)/2).$  The vertical dashed line marks the value of the constraint, $Y=F$; the horizontal dashed line mark the optimized risk $R(\optf).$    
\label{fig-2bivarLNcontoursandlossdstn} }
\end{figure}

\begin{figure}[p!]
\centering
\includegraphics[width=0.375\textwidth]{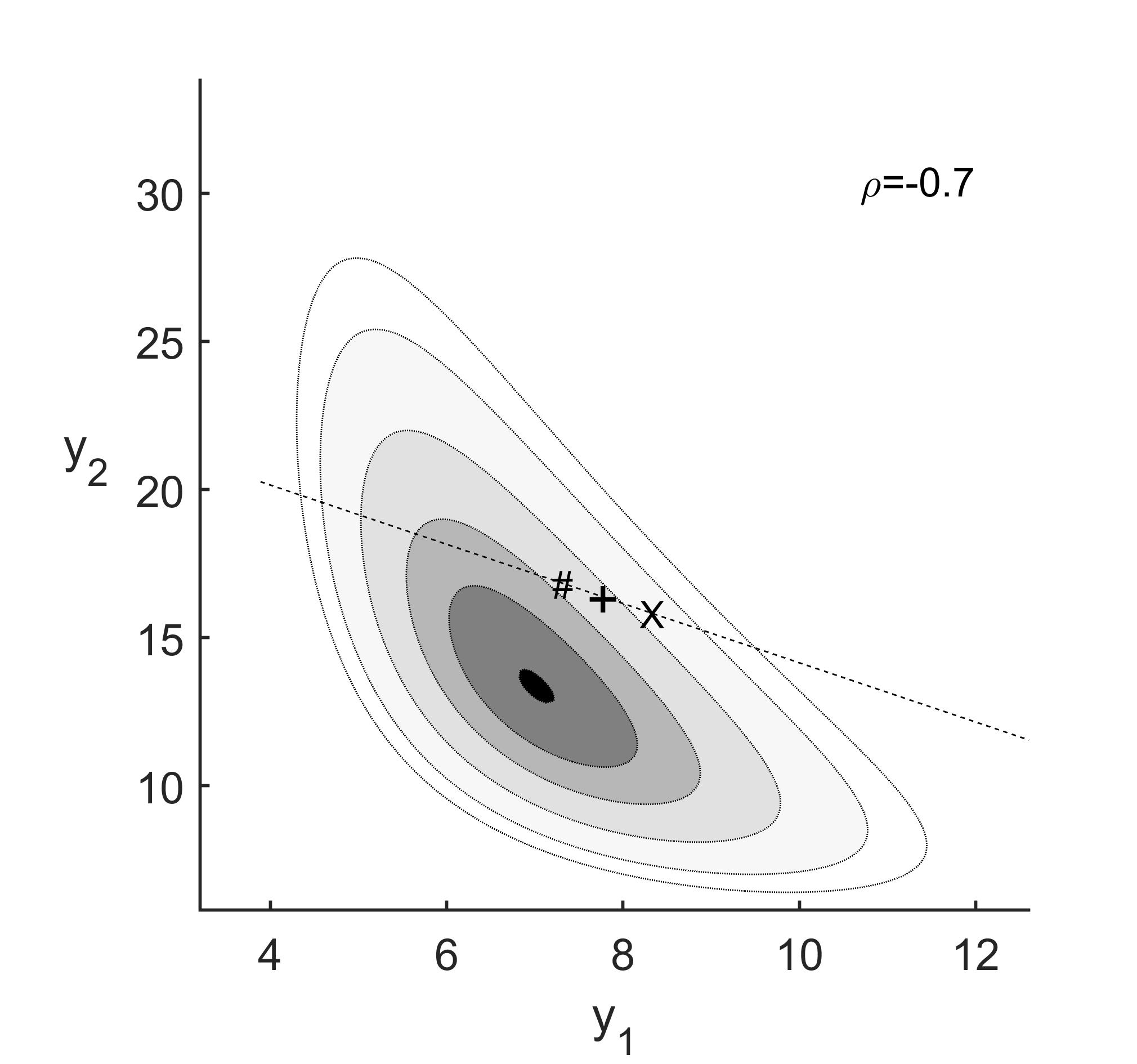}
\includegraphics[width=0.375\textwidth]{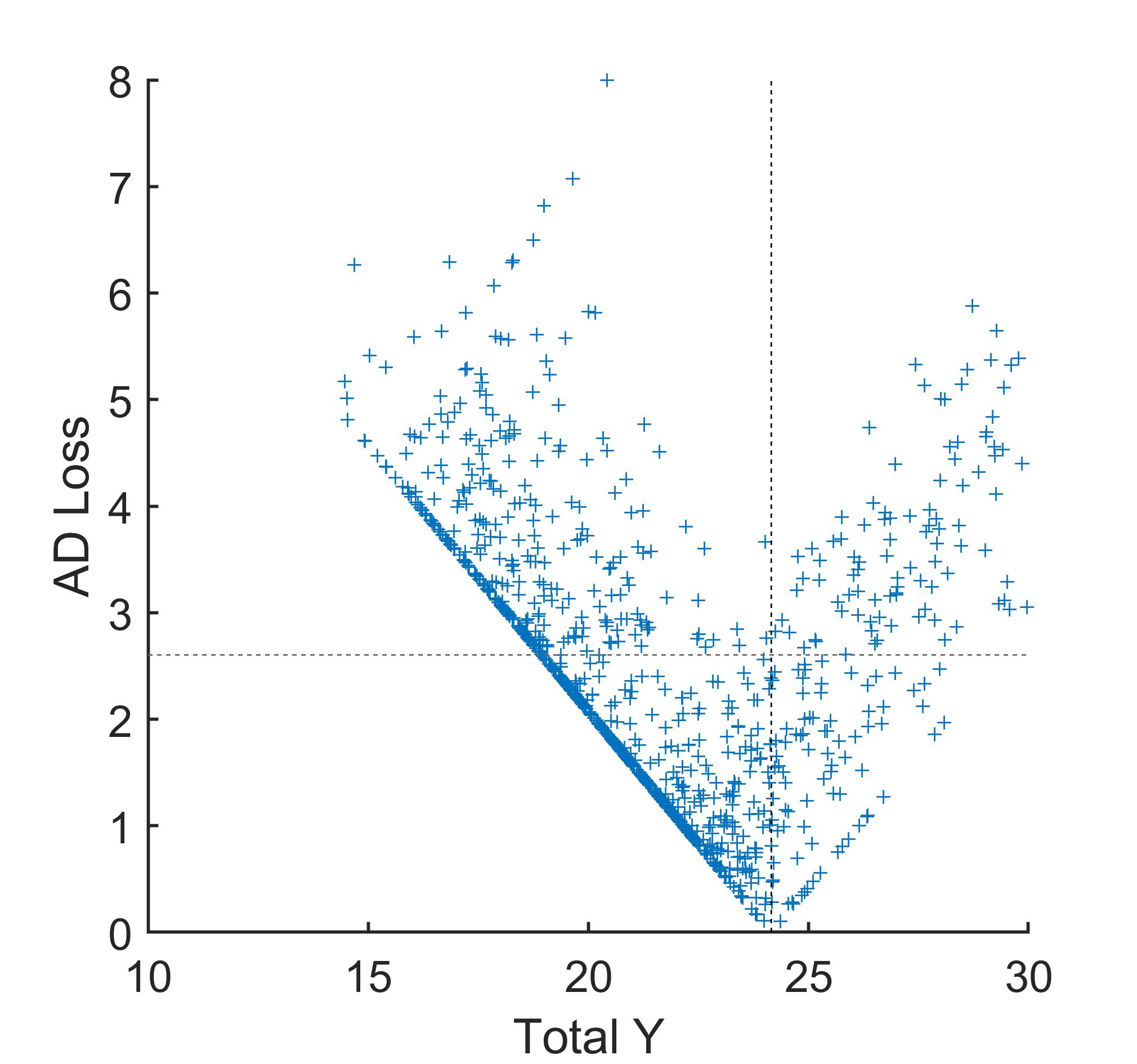}
\includegraphics[width=0.375\textwidth]{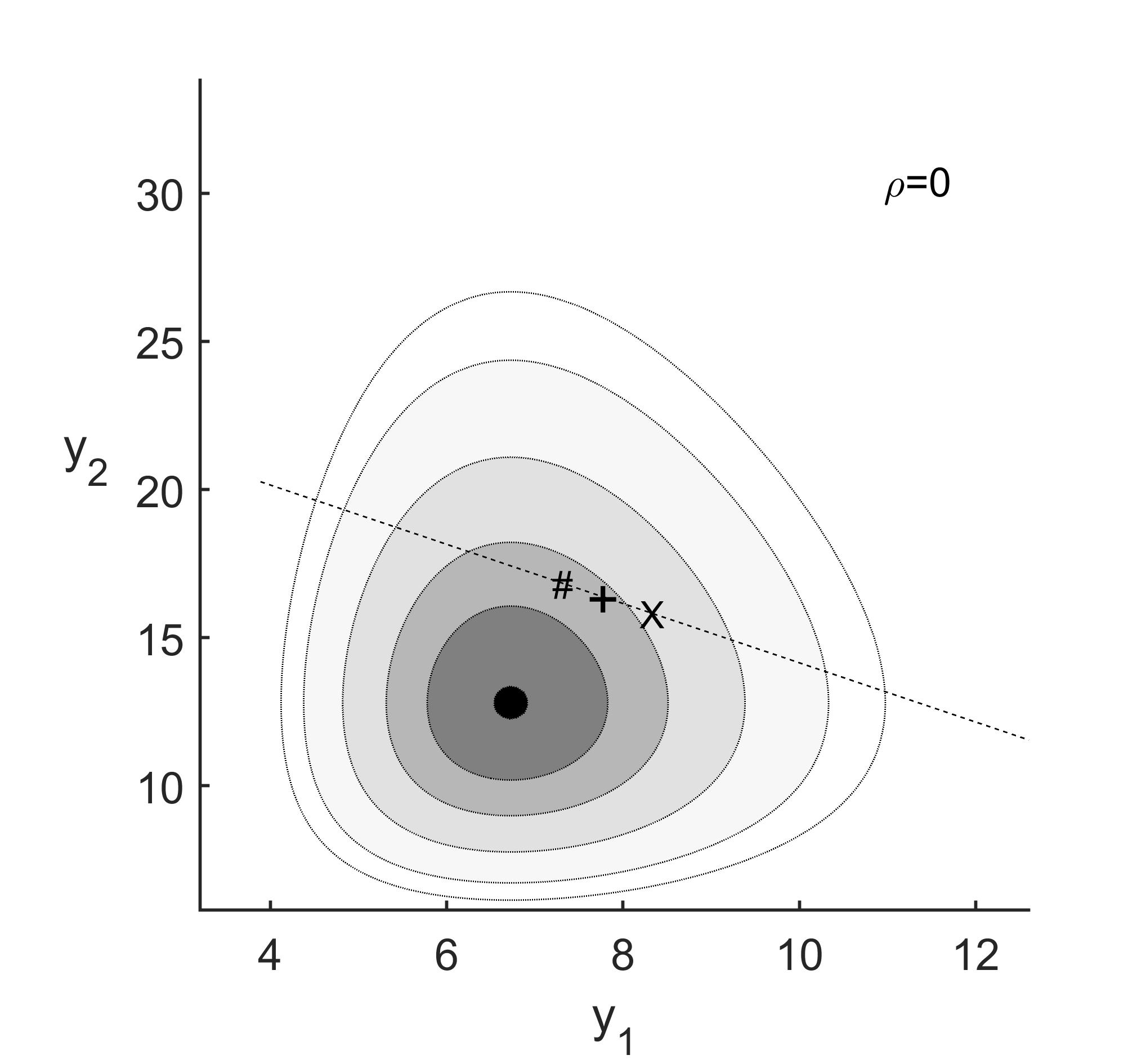}
\includegraphics[width=0.375\textwidth]{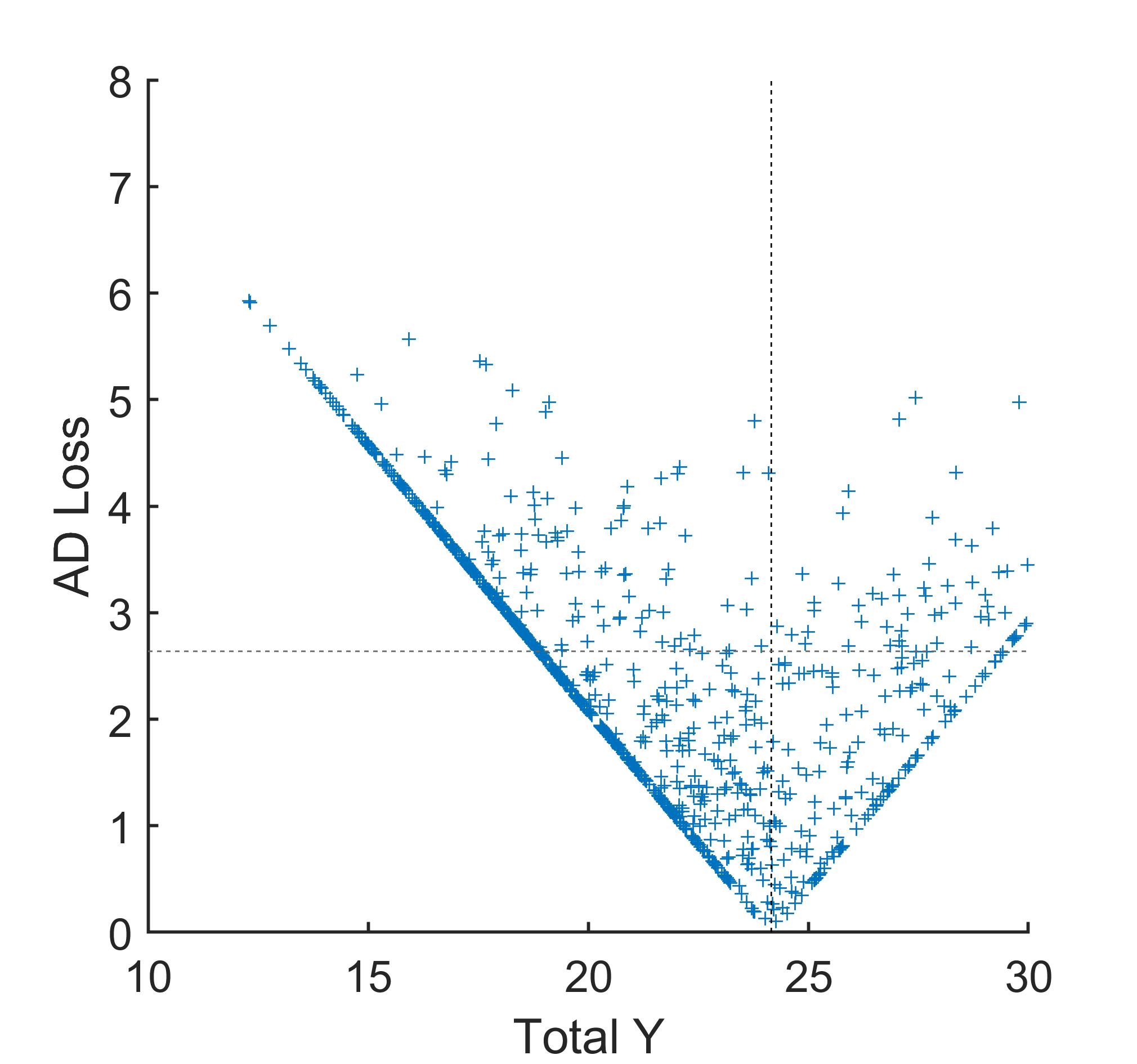}
\includegraphics[width=0.375\textwidth]{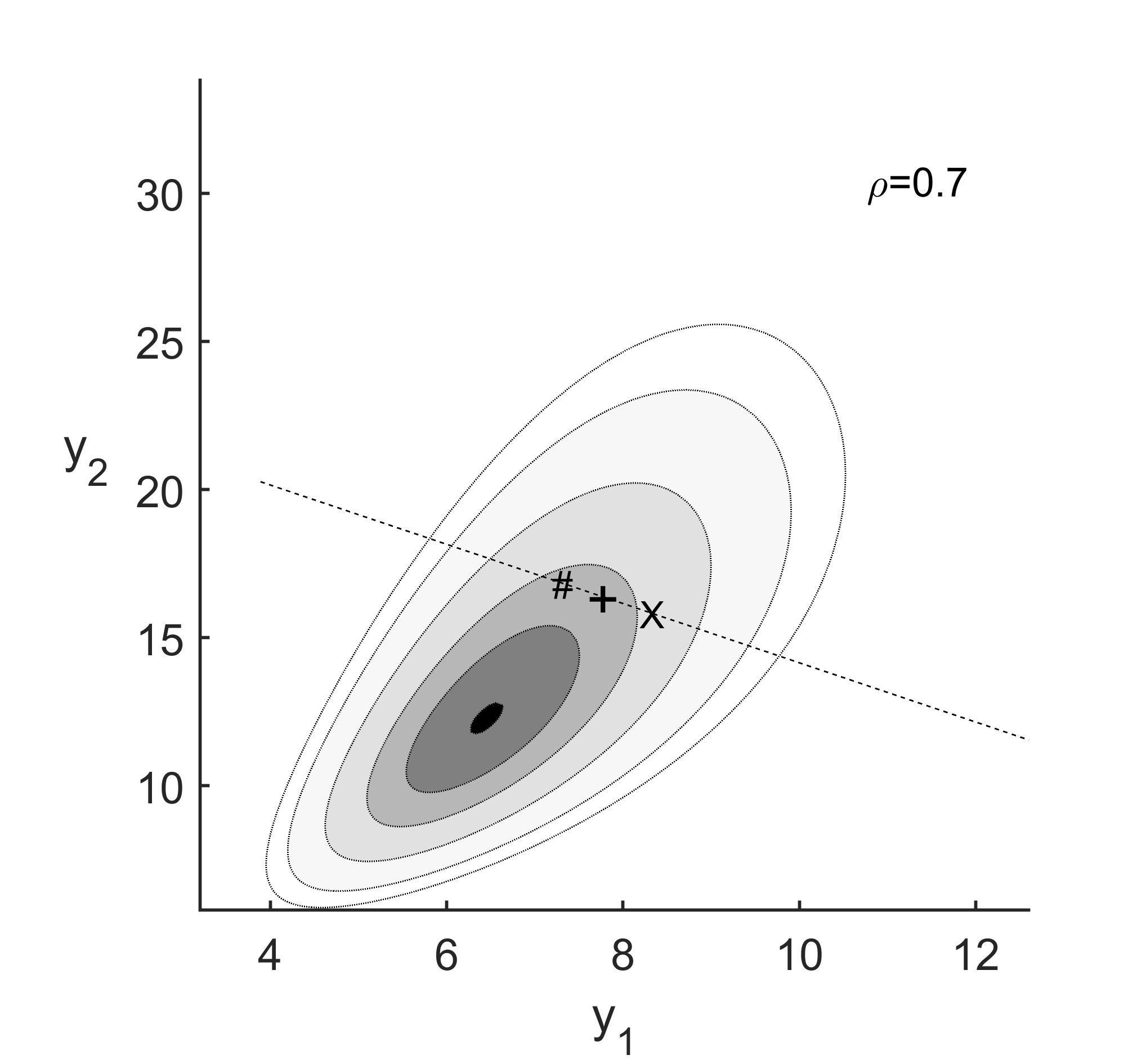}
\includegraphics[width=0.375\textwidth]{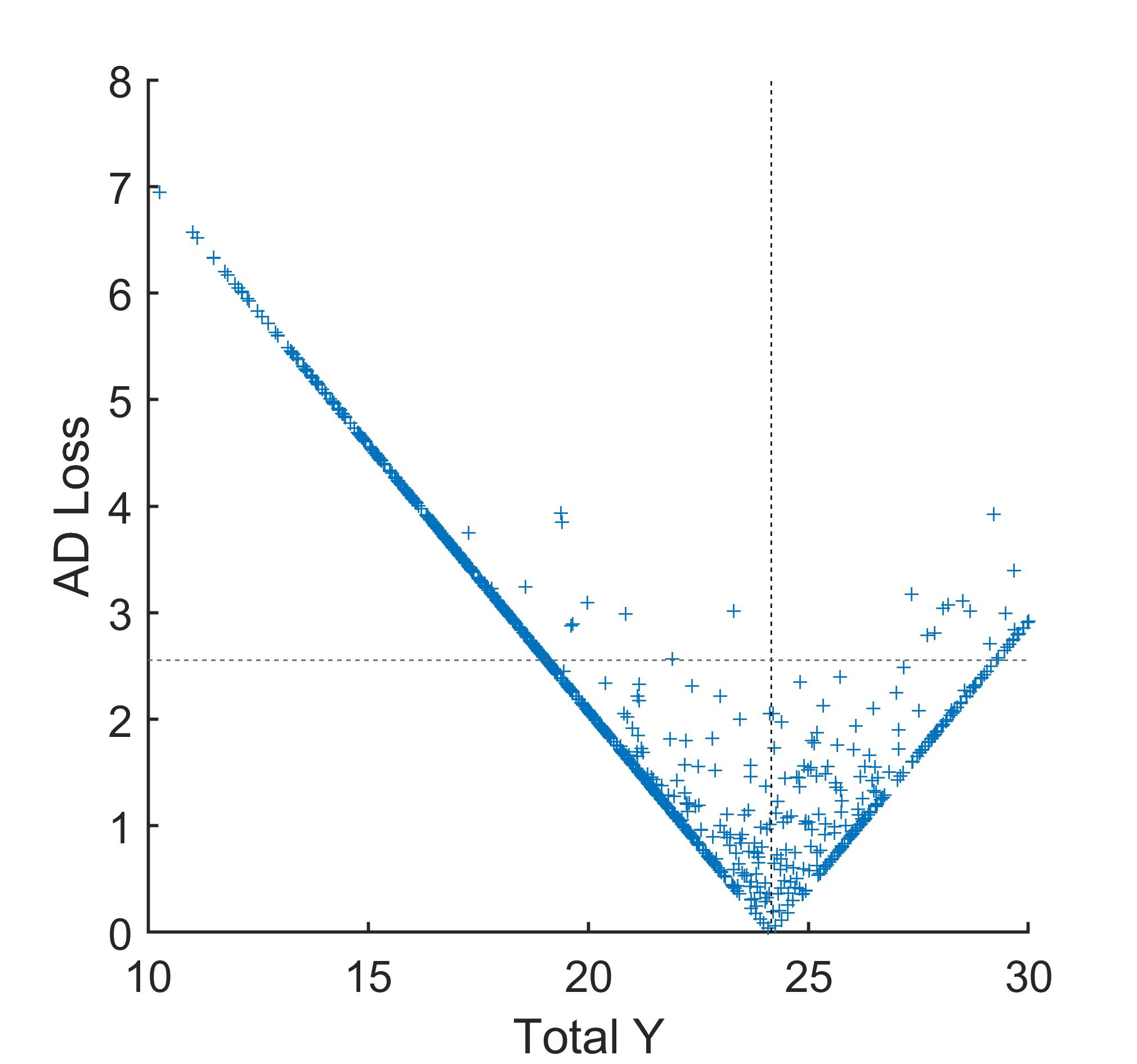}
\caption{Bivariate lognormal example as in Figure~\ref{fig-1bivarLNcontoursandlossdstn} of the text, but now with the value $F=24.15$ that lies in the upper tail of 
$P(Y).$  
{\em Left column:}  Contours of $p(\y)$ for three different levels of dependence $\rho\in \{-0.7,0,0.7\}$ and with the constraint $\one'\y=F$ indicated as the dashed line.  The symbols indicate the optimal point forecast vector $\optf$ under 
AD loss (+), APE loss (\#) and SE loss (X).   {\em Right column:}  Scatter plots of the corresponding   a Monte Carlo sample from $P(Y,L(\y,\optf)/2).$  The vertical dashed line marks the value of the constraint, $Y=F$; the horizontal dashed line mark  the optimized risk $R(\optf).$    
 \label{fig-3bivarLNcontoursandlossdstn} }
\end{figure}

Figures~\ref{fig-2bivarLNcontoursandlossdstn}~and~\ref{fig-3bivarLNcontoursandlossdstn} summarize two additional examples in the bivariate lognormal setting of 
Section~\ref{sec:bivarLNexamples}.  The example of that section exhibits results when the constrained total value $F=\one'\y$ is in the lower tail of its predictive distribution $P(Y)$.   The additional figures here show the same summaries under two other values:  a value very consistent with $P(Y),$ and a value well into the upper tail of $P(Y).$   In the first of these cases,  optimal forecasts of $\y$ conditioned on $F$ are very similar under AD, APE and SE losses, all being close to the center of the bivariate lognormal distribution.  The summaries in the second case parallel those of Section~\ref{sec:bivarLNexamples}, though the conditioning value $F$ is not so extreme as in the main text example.

\end{document}